\documentclass[useAMS,usegraphicx,usenatbib]{mn2e}

\def\ltsima{$\; \buildrel < \over \sim \;$}
\def\simlt{\lower.5ex\hbox{\ltsima}}
\def\gtsima{$\; \buildrel > \over \sim \;$}
\def\simgt{\lower.5ex\hbox{\gtsima}}
\makeatletter
\newcommand\ion[2]{#1$\;${\small\rmfamily\@Roman{#2}}\relax}%
\makeatother

\def\AA{$\; \buildrel \circ \over {\rm A}$}

\def\s{\ifmmode \widetilde \else \~\fi}
\def\={\overline}

\def\spose#1{\hbox to 0pt{#1\hss}}

\def\lta{\mathrel{\spose{\lower 3pt\hbox{$\mathchar"218$}}
     \raise 2.0pt\hbox{$\mathchar"13C$}}}
\def\gta{\mathrel{\spose{\lower 3pt\hbox{$\mathchar"218$}}
     \raise 2.0pt\hbox{$\mathchar"13E$}}}
\def\Dt{\spose{\raise 1.5ex\hbox{\hskip3pt$\mathchar"201$}}}    
\def\dt{\spose{\raise 1.0ex\hbox{\hskip2pt$\mathchar"201$}}}    

\def\dotsfill{\leaders\hbox to 1em{\hss.\hss}\hfill}

\title{The Galaxy Population of Abell 1367: The Stellar Mass-Metallicity 
Relation}
        
\author[Mouhcine et al.]{M.~Mouhcine$^1$, W.~Kriwattanawong$^{1,2}$,  
P.~A.~James$^1$\\
$^{1}$Astrophysics Research Institute, Liverpool John 
      Moores University, Twelve Quays House, Egerton 
      Wharf, Birkenhead, CH41 1LD, UK \\
$^{2}$Department of Physics and Materials Science, Faculty of Science, 
      Chiang Mai University
}

\date{Accepted ?. Received ?; in original form ?}

\pagerange{\pageref{firstpage}--\pageref{lastpage}}
\pubyear{2007}

\begin{document}

\maketitle

\label{firstpage}

\begin{abstract}

Using wide baseline broad-band photometry, we analyse the stellar population 
properties of a sample of 72 galaxies, spanning a wide range of stellar masses 
and morphological types, in the nearby spiral-rich and dynamically young galaxy 
cluster Abell 1367. The sample galaxies are distributed from the cluster centre 
out to approximately half the cluster Abell radius. The optical/near-infrared 
colours are compared with simple stellar population synthesis models from which 
the luminosity-weighted stellar population ages and metallicities are determined. 
The locus of the colours of elliptical galaxies traces a sequence of varying 
metallicity at a narrow range of luminosity-weighted stellar ages. 
Lenticular galaxies in the red sequence, however, exhibit a substantial spread 
of luminosity-weighted stellar metallicities and ages. For red sequence lenticular 
galaxies and blue cloud galaxies, low mass galaxies tend to be on average 
dominated by stellar populations of younger luminosity-weighted ages. 
Sample galaxies exhibit a strong correlation between integrated stellar mass 
and luminosity-weighted stellar metallicity. Galaxies with signs of morphological 
disturbance and ongoing star formation activity, tend to be underabundant 
with respect to passive galaxies in the red sequence of comparable stellar masses. 
We argue that this could be due to tidally-driven gas flows toward the star-forming 
regions, carrying less enriched gas and diluting the pre-existing gas to produce 
younger stellar populations with lower metallicities than would be obtained prior 
to the interaction. Finally, we find no statistically significant evidence for 
changes in the luminosity-weighted ages and metallicities for either red sequence 
or blue cloud galaxies, at fixed stellar mass, with location within the cluster.

\end{abstract}

\begin{keywords}
galaxies: formation -- galaxies: stellar content 
\end{keywords}

\section{Introduction}
\label{intro}

\footnotetext[1]{We dedicate this work to the memory of our friend and colleague 
C. Moss who died suddenly recently.}

Clusters of galaxies are unique laboratories for studying the impact of the
environment on galaxy properties. A large body of observational evidence 
has been collected over decades testifying that the properties and evolution 
of galaxies are strongly linked to their environment. The populations of 
galaxies in clusters tend to have spheroid-dominated morphologies
\cite[e.g.][]{dressler80,dressler97} with little or no ongoing star 
formation \cite[e.g.][]{bo84,lewis02}, in striking 
contrast with the properties of the dominant galaxy population in the field. 
What is unknown, however, is which dominant mechanisms cause this dichotomy, 
and in what kind of environments they mainly operate.

The hierarchical formation scenario predicts that galaxy clusters are formed
by accretion of building blocks of smaller mass at the nodes of large-scale
filaments \citep{west91,kw93}. 
The spiral-rich cluster Abell 1367 ($z\sim 0.0216$) lies at the intersection 
of two filaments in the Great Wall. The complex dynamical state of the 
cluster points towards the presence of multiple substructures falling into 
the core \cite[e.g.][]{cortese04} suggesting a comparatively recent 
formation of the cluster. These substructures contain a higher fraction 
of star forming galaxies than the cluster core, as expected during a merger 
event. This galaxy cluster is a rather unusual example of local, rich, 
dynamically young clusters since, in addition to massive evolved substructures, 
it is also experiencing the merging of a compact group directly falling into 
its core as revealed by deep narrow-band imaging, representing the region 
with the highest density of star forming systems ever observed in the local 
Universe \citep{sakai02,cortese04}.  X-ray observations of 
Abell 1367 show multiple clumps, extended gas features and a strong localised 
shock in the intra-cluster medium \citep{grebenev95,donnelly98}, 
that could be associated with a merging component in the process of penetrating 
the cluster. More recent X-ray data indicate the presence of cool gas streaming 
into the cluster core \cite[e.g.][]{sm02}, supporting a multiple merger 
scenario. Optical and radio observations also suggest that this cluster is 
currently experiencing group infall into its centre \citep{gavazzi01, gavazzi03}. 
The cluster has a high fraction of blue star-forming galaxies, which are more 
likely to appear disturbed and show evidence of tidal interaction driven star 
formation than similar galaxies in the field \citep{moss98,mw00,mw05}. In addition, 
galaxies exhibit the effects of 
ram-pressure of the inter-galactic medium \cite[IGM;][]{gavazzi01}.

The ages and the chemical properties of the stellar contents of galaxies 
are direct tracers of their star formation and chemical enrichment histories.  
Determining the properties of the current stellar population in galaxies allows 
the possibility of constraining the formation histories.  Clues to the properties 
of the stellar content, i.e., ages, metallicities, and abundance ratios, of
unresolved objects may be inferred from their integrated energy distribution. 
The problem is complicated, however, as the energy distributions of integrated 
stellar populations respond to variations of different parameters in degenerate 
ways. The most known of those is the so-called age-metallicity degeneracy 
\cite[e.g.][]{faber73,rose85,worthey94}, whereby optical broad-band colours 
and most absorption line strengths respond similarly to changes in age and 
metallicity. Major improvement was brought by the introduction of the Lick/IDS
system \citep{burstein84} and its extension to high-order Balmer lines
\citep{wo97}, which popularized the measurement of absorption 
line strengths in the spectra of galaxies. The development of models to predict 
the strengths of absorption indices as a function of stellar population parameters 
\cite[e.g.][]{worthey94,bressan94,vazdekis99,thomas03,schiavon07} has shown that 
metallic lines and Balmer lines show different sensitivities to ages and metallicities, 
thus allowing the age-metallicity degeneracy to be broken.

A complementary approach to derive constraints on stellar population ages and 
metallicities is the use of optical/near-infrared colours 
\cite[e.g.][]{james06, lee07}. A number of observational studies highlight the 
potential of the combination of optical and near-infrared (hereafter near-IR) 
integrated colours for breaking the age-metallicity degeneracy for both globular 
clusters \cite[e.g.][]{puzia02, hk04}, and stellar systems with complex star 
formation histories \cite[e.g.][]{peletier90,pb96,smail01,macarthur04}.
\citet{james06} have used the suite of models used in the present paper to 
investigate the extent to which the stellar 
population parameters derived from broad-band optical/near-IR agree with other 
constraints derived independently. They found that the colours of the Large 
Magellanic Cloud and M31 globular clusters are consistent with their predicted 
colours given their luminosity-weighted ages and metallicities available in the 
literature. The mean ages derived from the integrated colours of two Local Group 
dwarf galaxies are consistent with the star formation histories inferred from their 
resolved stellar populations. 

A number of studies have compared properties of galaxies in clusters 
and in the field to constrain the environmental effects on galaxy properties 
\cite[e.g.][]{rose94,tf02,thomas05,sanchez-blazquez06}. A complementary approach 
to investigate the environmental processes is to compare the properties of 
galaxies at different radii from the cluster centre 
\cite[e.g.][]{guzman92,carter02,smith06,smith08}. In hierarchical clustering 
simulations, the average accretion epoch of subhaloes in cluster-sized 
parent haloes appears to be correlated with their final radius from the 
cluster centre \citep{gao04}. Gao and his collaborators have found 
that galaxies within 20 per cent of the virial radius at the present epoch 
were accreted $\sim8$\,Gyr ago, while those at the virial radius entered 
the cluster only $\sim4$\,Gyr ago. Independently of the precise nature of 
physical mechanism(s) that suppress star formation in the cluster environment 
and the timescales involved, the gradual accretion of galaxies into the cluster 
would likely lead into cluster-centric gradients in the stellar population properties 
is expected. 

In the present paper, we will examine the properties of galaxies in the nearby 
cluster Abell 1367 by determining the luminosity-weighted ages and metallicities 
of the dominant stellar populations of Abell 1367 galaxies, and the cluster-centric 
gradient of galaxy properties. The layout of the paper is as follows. 
In section \ref{data}, we first describe briefly the data set employed and the 
selection of the sample of the cluster members, while in section \ref{stellar_pop} 
we present the relationships between the stellar mass and luminosity-weighted 
stellar population properties of our sample, and determine the radial trends 
followed by the properties of the cluster galaxies. Finally, our principal conclusions 
are represented in section \ref{summary}.

\section{Data and sample selection}
\label{data}

The observations, the data reduction techniques applied to the multi-wavelength 
data set used in the present paper and the selection of the final galaxy sample 
will be described elsewhere. We therefore only briefly summarize this information 
here.

Our photometric and spectroscopic observations of the Abell 1367 cluster were 
taken from a number of facilities. UBR broad-band imaging and narrow-band imaging 
observations over three pointings of the Wide Field Camera on the INT, 
i.e., $34\times90$ arcminute square, were obtained. The exposure times on U-band 
and B-band filters were $3\times300$ and $3\times150$ seconds respectively. 
The R-band images were taken with an exposure time of 300 seconds, whereas the 
narrow-band [SII] filter images were obtained for $3\times400$ seconds. 
This narrow-band filter was used as it samples the H$\alpha$ emission at the 
distance of Abell 1367. The narrow-band filter used includes both H$\alpha$ and 
the adjacent [NII]$\lambda6548,6583$ emission lines. For the rest of the paper, 
we will refer to this only as H$\alpha$. J and K broad-band imaging data were 
obtained using the Wide Field Camera on UKIRT, covering 0.75 square degrees, 
centered on the cluster centre. 

All photometric data were reduced by the Cambridge Astronomical Survey Unit, with 
standard pipelines for INT/WFC and UKIRT/WFCAM imaging data. As there are not many 
standard stars for zero-point calibration in the observed fields, we have used 
published galaxy photometry as photometric standards. Magnitudes quoted in specific 
apertures were preferred for this calibration. The zero-points in INT/WFC U-band 
and B-band photometry used apertures sizes given in \citet{buta96}. 
The zero-point for the R-band photometry was calibrated using the photometric data 
from \citet{vl98} and \citet{taylor05}. Near-IR imaging data 
were calibrated using the Two Micron All Sky Survey photometry. 
Photometry was corrected for foreground Galactic extinction using the reddening 
maps of \citet{schlegel98}. We have not attempted to correct the photometry 
for internal extinction. K-correction, as determined by \citet{poggianti97}, was applied 
for all images in all filters. The INT/WFC B-band images were used to visually classify 
the morphologies of the galaxies in our sample.  Signs of disturbance were defined 
for the sample galaxies that have clear structures, but show i.e. asymmetries, 
bridges/tails, broken spiral arms, or merger-like morphologies.

To select a sample of likely members of the cluster, we have proceeded as follows. 
The sample selection started from H$\alpha$ detections within the central area of 
the cluster. 86 emission line objects with signal-to-noise larger than five are 
detected. In order to remove the contamination from foreground stars, objects with 
full width half maximum smaller than seven pixels, i.e., 5$\sigma$ above the stellar 
mean, were excluded. The emission line sample was used to estimate the B-band 
magnitude limit of galaxies for which H$\alpha$ emission could be detected. 
Objects in this sample are all brighter than B=21 magnitude in B-band. This magnitude 
was adopted as the faint limit of our sample. These criteria were applied to all 
extended objects 

In order to minimise the contamination from background objects, the surface number 
densities of galaxies were used to estimate the cluster radius and the probable 
fraction of cluster members. The surface number density of objects brighter than 
21 and full width half maximum smaller than seven pixels, is found to drop around 
half of the Abell radius of the cluster, and these objects are distributed uniformly 
beyond this radius. Bright objects appear in higher proportion within this radius. 
We have then decided to consider only galaxies within half of the Abell radius. 
Then the selection of galaxies was limited by a diameter criterion to optimise 
the selection of true cluster members. The final selected sample of cluster galaxy 
candidates comprised 304 galaxies. For these galaxies, isophotal $R_{24}$ apertures 
were determined to measure UBRJK magnitudes and H$\alpha$ fluxes.

To determine the cluster membership for galaxies in the photometrically selected 
sample, spectroscopic observations from the WYFFOS fibre-fed spectrograph on the 
William Herschel Telescope were obtained to determine their redshifts. 
The total area covered by WYFFOS/WHT observations covers approximately the same 
field as the INT/WFC imaging observations, with exposure times of 
$4\times1800$ seconds for four pointings and $3\times600$ seconds for one pointing. 
The grating used gives a spectral dispersion of approximately 3\,\AA ~pixel$^{-1}$, 
with a total coverage of 3000\,\AA, centred on 6000\,\AA, covering a number of 
absorption and emission lines. The observations targeted preferentially faint, blue, 
and small angular size galaxies in the photometrically selected sample. The faintest 
object has a total B-band magnitude of 22, and most of the targeted objects are 
brighter than 21.5 in R-band. The spectroscopic data were reduced using the WYFRED 
package, a special pipeline for the reduction of multifibre spectroscopy obtained 
by WHT/WYFFOS. Bias-substracted, median-combined science frames for each pointing, 
and arcs and sky images used as input for the WYFRED package. The redshifts measured 
from several lines were averaged for each galaxy. Among the 304 galaxies selected 
from broad-band photometry data, 128 galaxies have measured redshifts. 
72 among them are associated with the Abell 1367 cluster, whereas 56 are background 
objects. The new cluster members have velocities within 2.5 times the velocity 
dispersion from the average velocity of the cluster as determined by \citet{sr99}. 
Most of galaxies with spectroscopically determined membership lie within approximately 
half of the Abell radius of the cluster. The B-band magnitudes of the cluster members 
range from 16 to 20, with all galaxies brighter than 16 magnitudes in the 
photometrically selected sample being cluster members. The fraction of cluster members 
to the known redshift galaxies decreases dramatically from $\sim80\%$ for galaxies 
with B-band magnitude ranging from 16 to 17, to 10\% for galaxies with magnitude 
between 19 and 20.

The observed properties of the final sample of Abell 1367 used in the rest of the paper 
are given in the first five columns of Table~\ref{gal_prop_obs1}, giving the galaxy 
name (column 1), the morphological type (column 2), the presence of morphological 
disturbance (column 3; y: yes, n: no),  and optical/near-IR colours (column 4 and 5).

\section{Stellar population properties of Abell 1367 galaxies}
\label{stellar_pop}

\subsection{Colour-magnitude and colour-colour diagrams}
\label{col_col}

\begin{figure}
\includegraphics[clip=,width=0.5\textwidth]{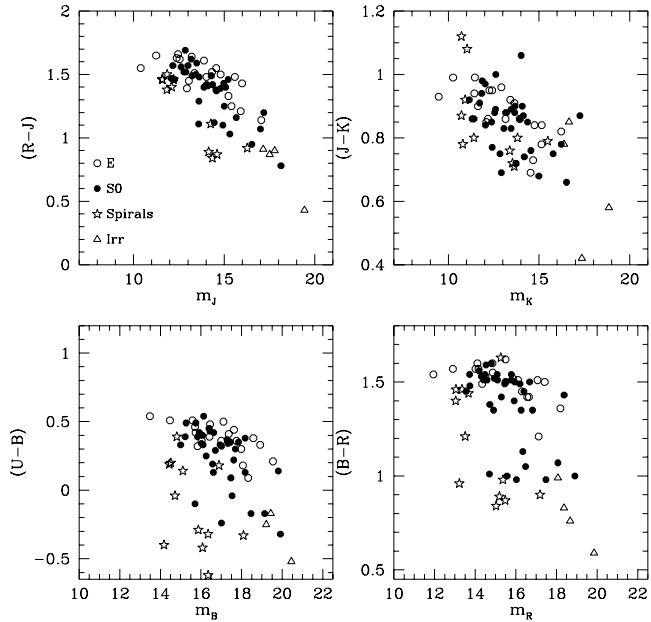}
\caption{Colour-magnitude relations for our sample with galaxies separated by 
morphological type.}
\label{cmr_morph}
\end{figure}

\begin{figure}
\includegraphics[clip=,width=0.5\textwidth]{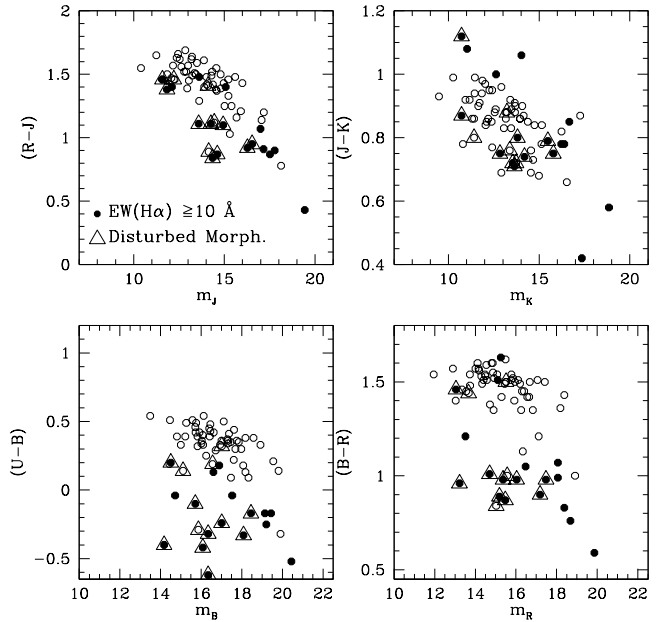}
\caption{Similar to figure \ref{cmr_morph}, but galaxies are separated by H$\alpha$ 
emission line equivalent width and the presence of signs of disturbance.}
\label{cmr_ew}
\end{figure}

Figures \ref{cmr_morph} and \ref{cmr_ew} show the distribution of the galaxies in our 
sample in four optical/near-IR colour-magnitude diagrams. The galaxies in both figures 
are coded on the basis of their morphological type, equivalent width of the H$\alpha$ 
emission line and the presence of morphological disturbance. These illustrate the wide 
range in luminosity covered in our analysis, nearly six magnitudes, with galaxies with 
different properties spread across this range, except the brightest two galaxies which 
are ellipticals. The colour-magnitude diagrams of our sample galaxies show the normal 
bimodal structure, in Abell 1367 the blue sequence of late-type galaxies is clearly 
visible, in agreement with previous studies of this cluster \citep{gp82}, 
but in contrast to the colour-magnitude diagrams of more relaxed clusters such as Coma 
\citet[e.g.][]{terlevich01}. Specifically \citet{terlevich01}
find that most spiral, irregular, and unclassified galaxies in their Coma sample lie 
on the same (U-V) colour-magnitude relation as the early-type galaxies, with a small 
fraction populating the blue sequence. In contrast, for our Abell 1367 sample, only 
two of the 17 galaxies with late-type morphologies have colours consistent with the 
red sequence, one of which has an H$\alpha$ emission line equivalent width larger than 
10\,\AA. The bimodality is less clear in the near-IR colour-magnitude diagrams as the 
near-IR is less sensitive to recent and ongoing star formation.

Interestingly, a number of lenticular galaxies in our sample show H$\alpha$ emission 
and blue optical/near-IR colours, and exhibit signs of morphological disturbance, 
belonging thus to the blue cloud. 

\begin{figure*}
\includegraphics[clip=,width=0.45\textwidth]{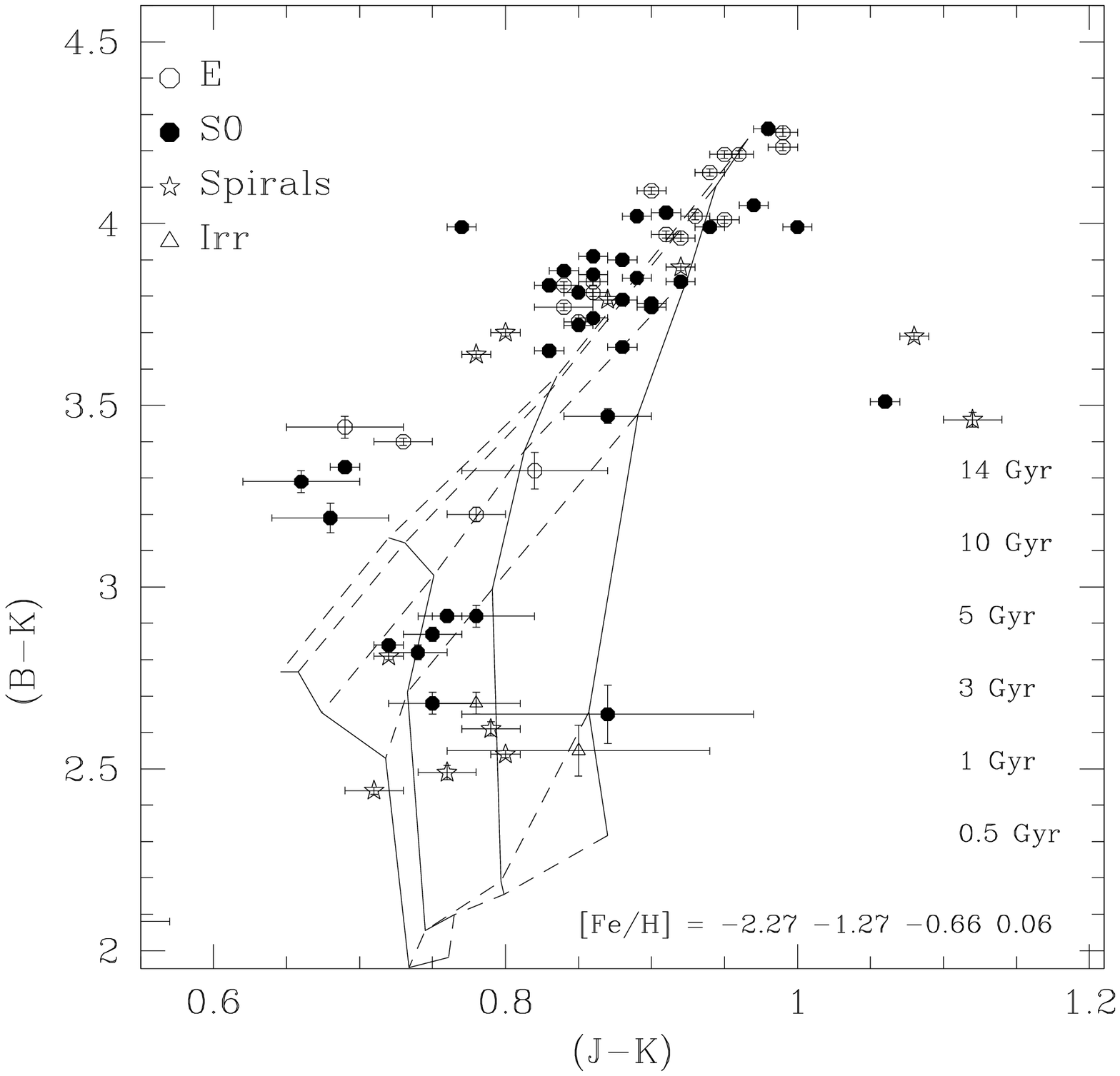}
\includegraphics[clip=,width=0.45\textwidth]{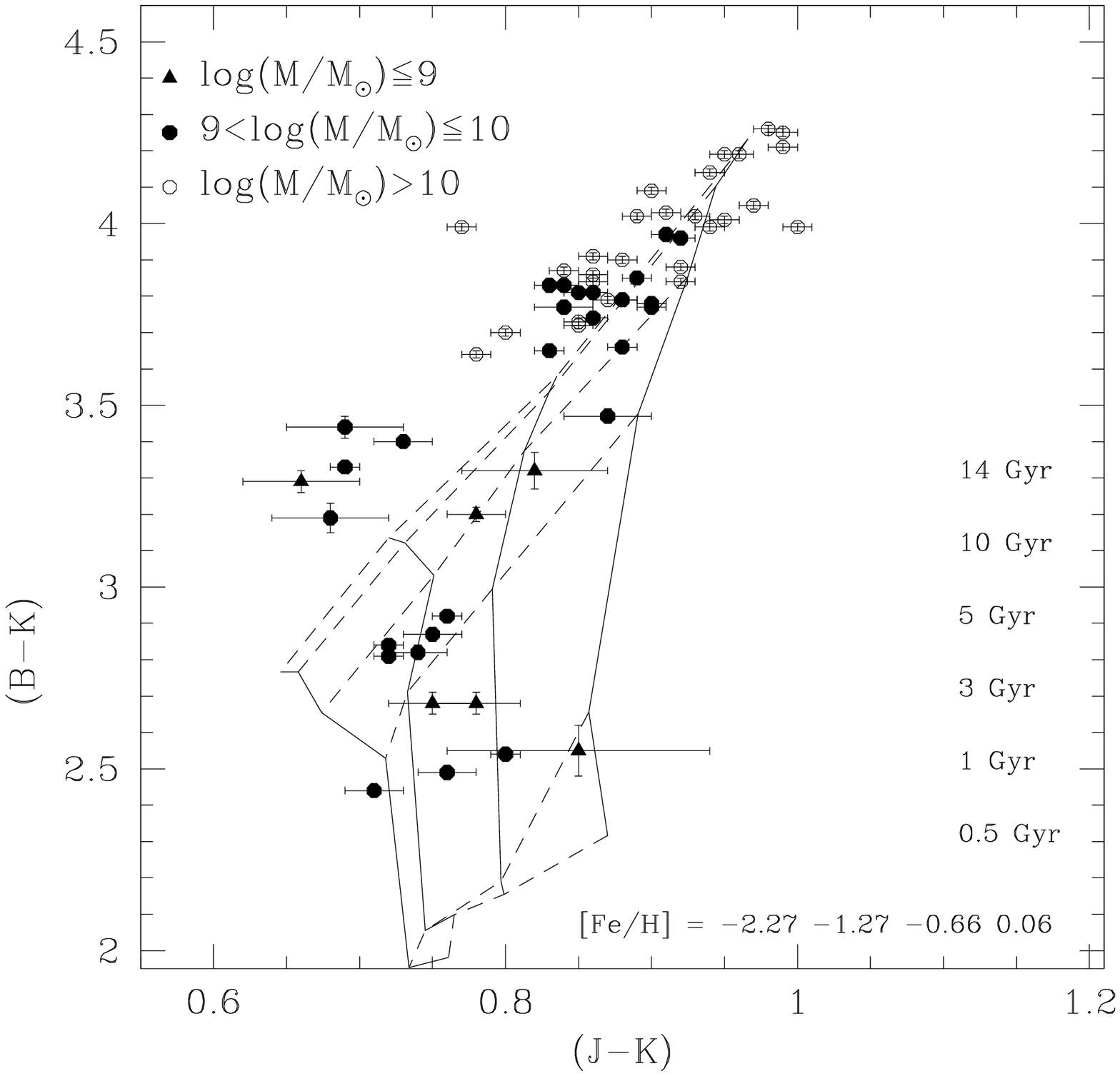}
\includegraphics[clip=,width=0.45\textwidth]{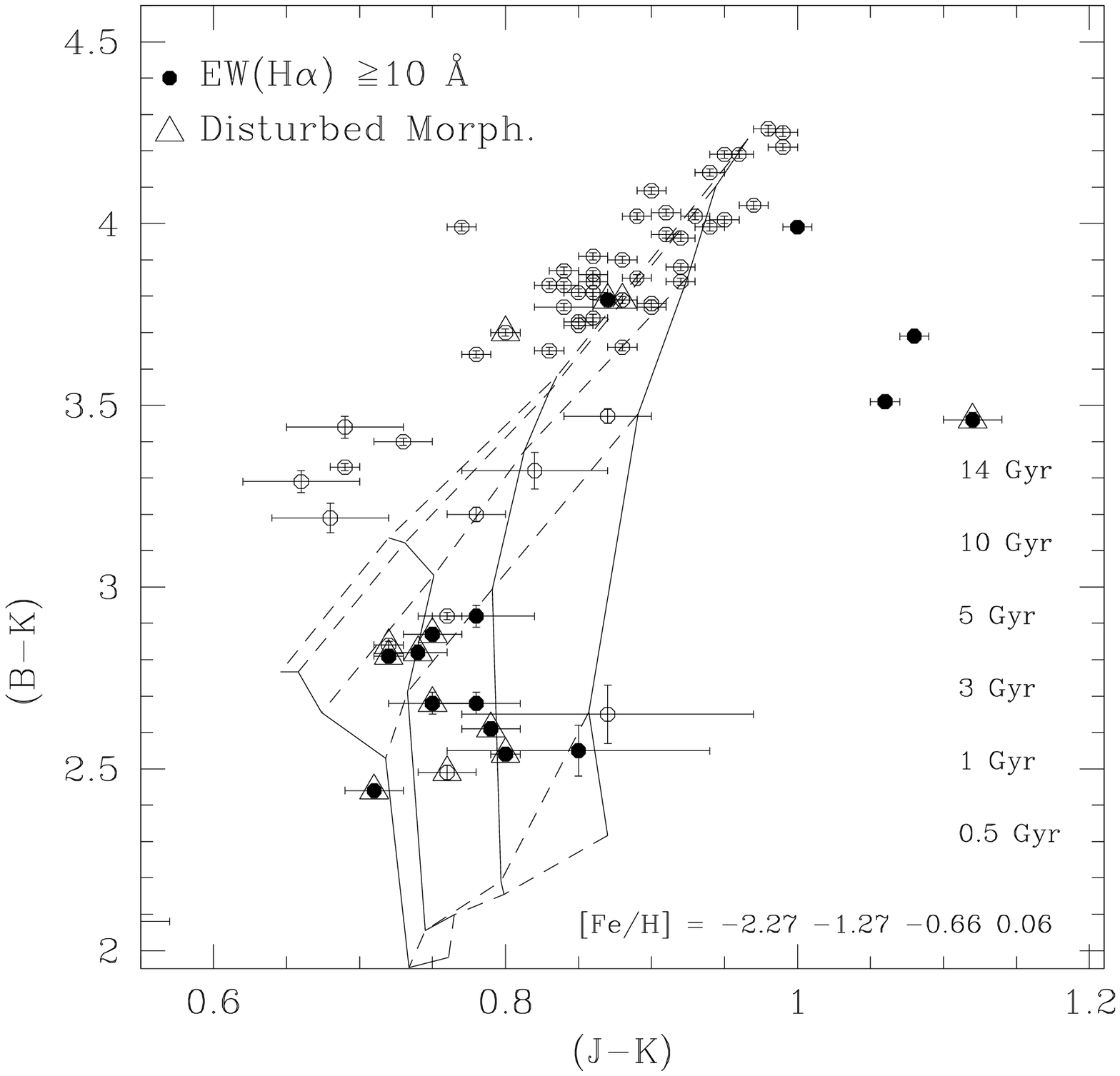}
\includegraphics[clip=,width=0.45\textwidth]{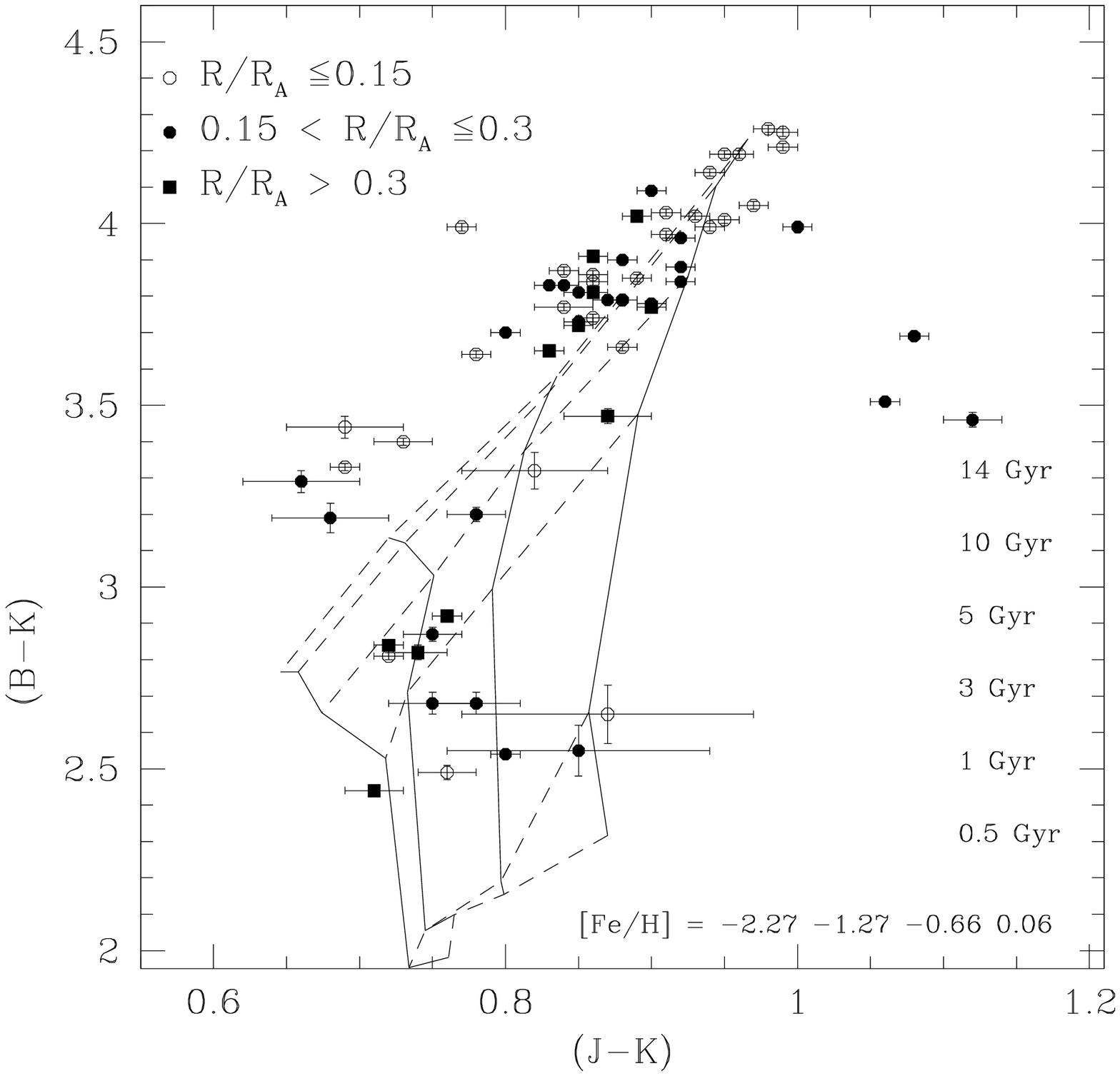}
\caption{Optical/near-IR colour-colour diagrams for the sample of galaxies studied in 
the present paper separated by morphological type (the upper-left panel), stellar mass 
(the upper-right panel), equivalent width of H$\alpha$ emission and the presence of 
morphological disturbance (the lower-left panel), and cluster-centric distance (the 
lower-right panel). Overplotted are the \citet{pietrinferni04} stellar population 
models for a simple stellar population. Solid and dashed lines connect models of 
constant metallicities and ages, respectively. }
\label{bmk_jmk}
\end{figure*}

Figure \ref{bmk_jmk} shows the distribution of our sample of Abell 1367 galaxies in 
the (B-K) vs. (J-K) colour-colour diagram, separated by morphological type (upper 
left panel), integrated stellar mass (upper right panel), equivalent width of H$\alpha$ 
emission and signs of the presence of morphological disturbance (lower left panel), 
and cluster-centric radial distance (lower right panel). In order to convert the 
photometric information into constraints on the underlying ages and metallicities 
of the stellar populations, the optical/near-IR photometry must be compared with 
predictions of stellar population synthesis models. In the simplest version of such 
models, commonly termed simple stellar populations, all stars are assumed to originate 
from a single star formation event and thus to have identical chemical compositions 
and ages. Theoretical predictions for such stellar populations from the models of 
\citet{pietrinferni04} are overplotted on Fig. \ref{bmk_jmk}. The grid of models is 
built assuming a \citet{kroupa01} initial mass function, and solar $\alpha$-to-iron 
ratio. The models span a wide range of ages, i.e., 0.5, 1., 3., 5, 10 and 14 Gyr, and 
initial metallicities, i.e., [Fe/H]=-2.27, -1.27, -0.66, and 0.06. Models of constant 
age are the solid lines, while models of constant metallicity are dotted lines. 
As clearly shown in the figure, the combination of optical and near-IR colours is 
able to disentangle the effects of age and metallicity. A comparison of the model 
grid and the observed colours of our sample shows good agreement between the 
regions of the colour-colour plane populated by the observations and those where 
galaxies are expected to lie on the basis of the stellar population synthesis models. 
By interpolating from the position on the optical/near-IR colour-colour diagram, it
is possible to derive the ages, metallicities of the dominant stellar populations, and 
their associated errors.

By comparing the location of the sample galaxies in the colour-colour diagram 
to predicted colours, a number of features emerge. The dominant feature in the 
colour-colour diagram is a tight locus of points in the grid region corresponding 
to the oldest and most metal rich stellar population models. These galaxies are 
typically the brightest elliptical and lenticular galaxies in the cluster. 
Their locus traces what appears to be a metallicity sequence, spanning a range 
of more than one dex in metallicity at a narrow age range, i.e., $\sim13$\,Gyr 
(although the uncertainties in the metallicities and the relative calibration of 
the colours and grids means that one has to view the ages as defined on a relative, 
rather than an absolute scale). There is a tendency for these red sequence galaxies 
to lie above the grid, with rather redder (B-K) colours than predicted by the models 
even for the oldest ages. One possible explanation for this is the effect of atomic 
diffusion in stellar interiors. Such gravitational settling of helium and heavier 
elements has been demonstrated to occur in the Sun \citep{guenther94} but its 
occurrence in old stellar populations such as globular clusters is uncertain 
\citep{gratton01,korn06}.  As a result, this effect is not included in the BaSTI 
models. If it were to be included, the model grid would extend to redder (B-K) 
colours for old stellar populations. 

In contrast with the elliptical galaxies, for which the low mass objects appear to 
extend the metallicity sequence defined by the brighter ellipticals to low stellar 
metallicities, the optical/near-IR colours of less massive lenticular galaxies 
show a different colour-colour relationship, consistent with the appearance of 
a population of lenticular galaxies with young luminosity-weighted mean ages and 
low stellar metallicities.

Another prominent feature in the colour-colour diagram is the presence of a population 
of galaxies with optical/near-IR colours consistent with those of synthetic stellar 
populations of young ages and low metallicities. None of the ellipticals belongs to this 
group of galaxies. All members of this population belong to the blue cloud as defined 
in the colour-magnitude diagrams (Fig. \ref{cmr_morph} and Fig. \ref{cmr_ew}). 
All are disturbed morphologically with ongoing star formation activity. Their stellar 
masses can be as high as those of galaxies in the tight metallicity sequence populated 
by old and passive red sequence galaxies, and extend down to lower masses.

A small number of galaxies in our sample show very red (J-K) colours, and lie well 
outside the grid of models. One of them is UGC 6697, a well studied system of what 
appears to be a merger of two edge-on galaxies \citep{gavazzi01}, accompanied by a 
vigorous star forming event with H$\alpha$ emission equivalent width of 94\,\AA. 
The geometry of the system suggests that the photometry could be severely affected 
by internal reddening. The second object is G1143498+195835, a low mass lenticular, 
which lies close to UGC 6697, with a projected distance of $\sim7$\,kpc. 
Despite the ongoing ram pressure gas stripping \citep{bg06}, there is apparently 
still a sufficient gas reservoir in this galaxy to sustain a star formation 
event, with a measured H$\alpha$ emission line equivalent width of 31\,\AA. 
The third object is UGC 6702, a bright spiral galaxy with circumnuclear star formation, 
and H$\alpha$ equivalent width of 32\,\AA. For the last two objects, the red 
(J-K) colour and the intermediate (B-K) colour could be interpreted as a possible 
signature of the presence of a sizeable population of red intermediate mass carbon 
stars. Additional near-IR spectroscopy could help understanding the optical/near-IR 
photometric properties of these objects.

\begin{figure*}
\includegraphics[clip=,width=0.45\textwidth]{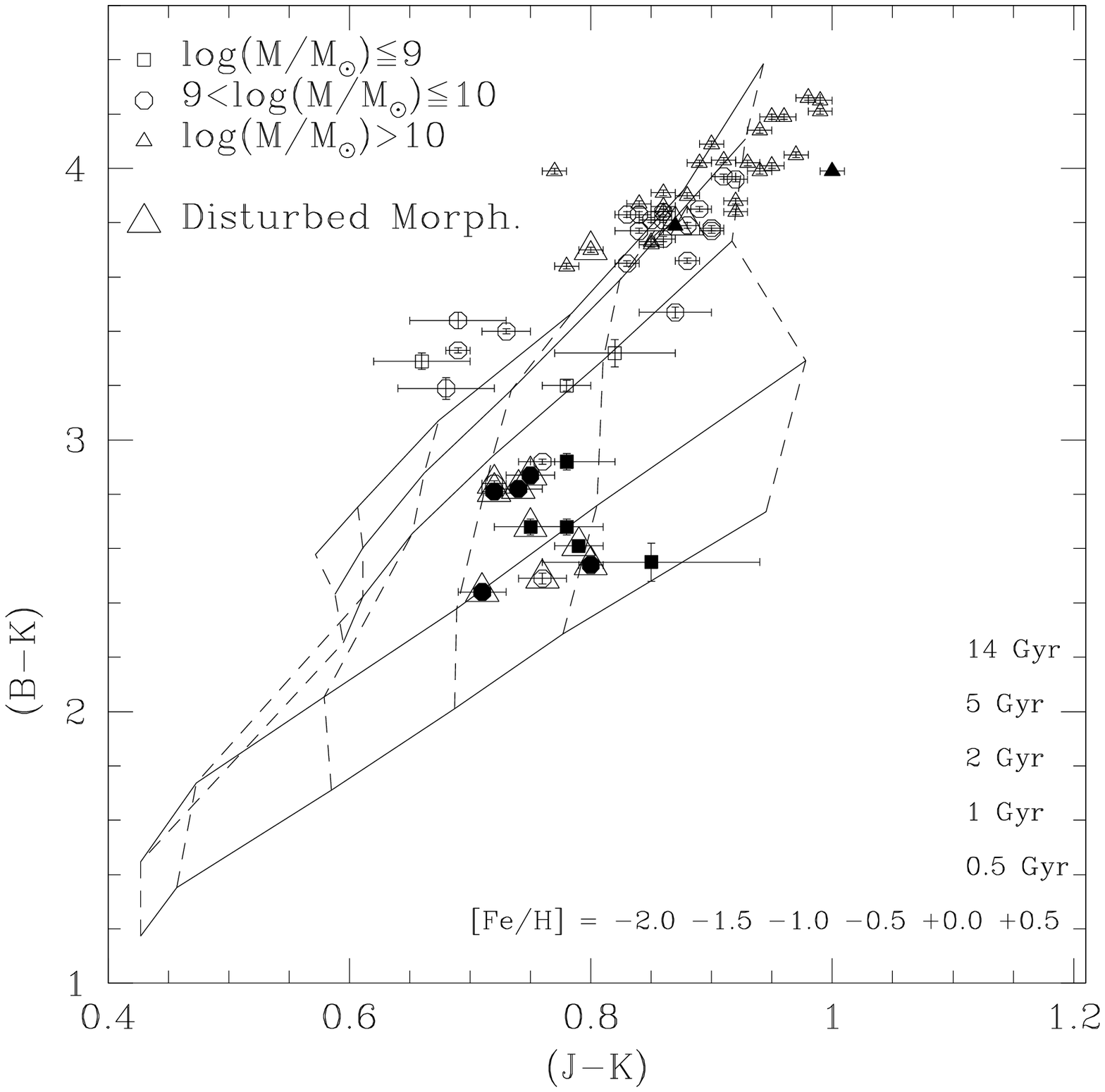}
\includegraphics[clip=,width=0.45\textwidth]{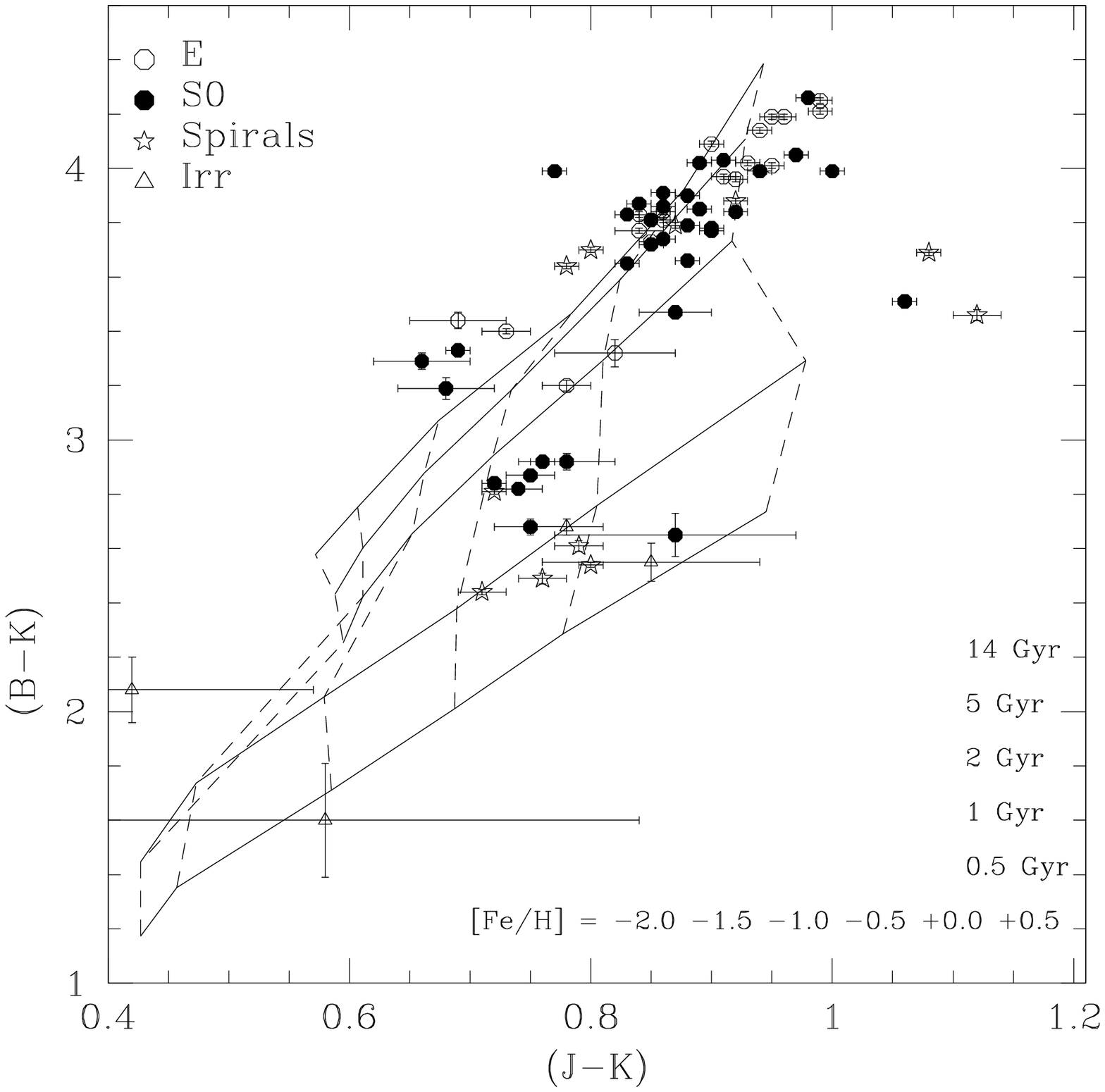}
\includegraphics[clip=,width=0.45\textwidth]{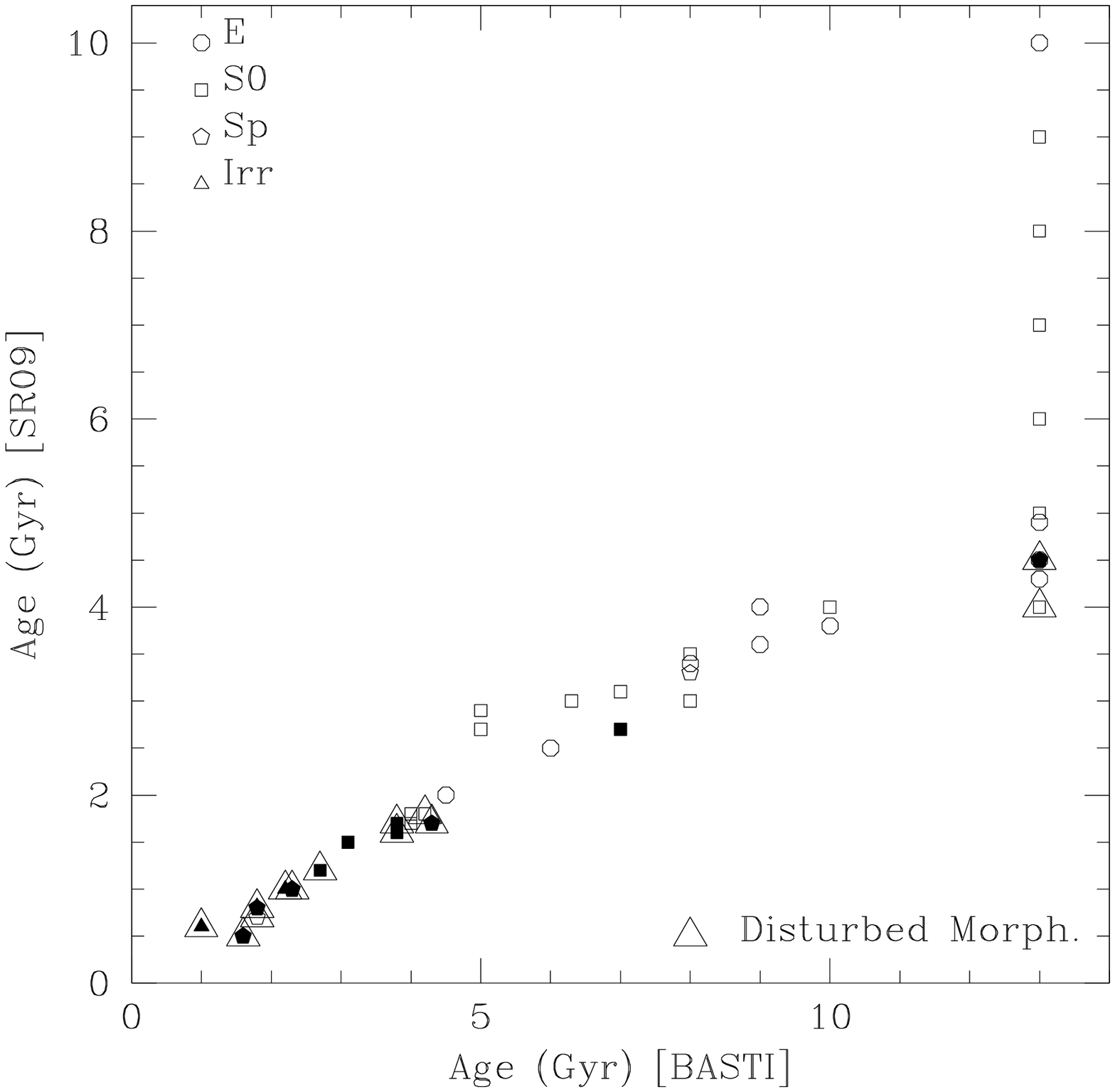}
\includegraphics[clip=,width=0.45\textwidth]{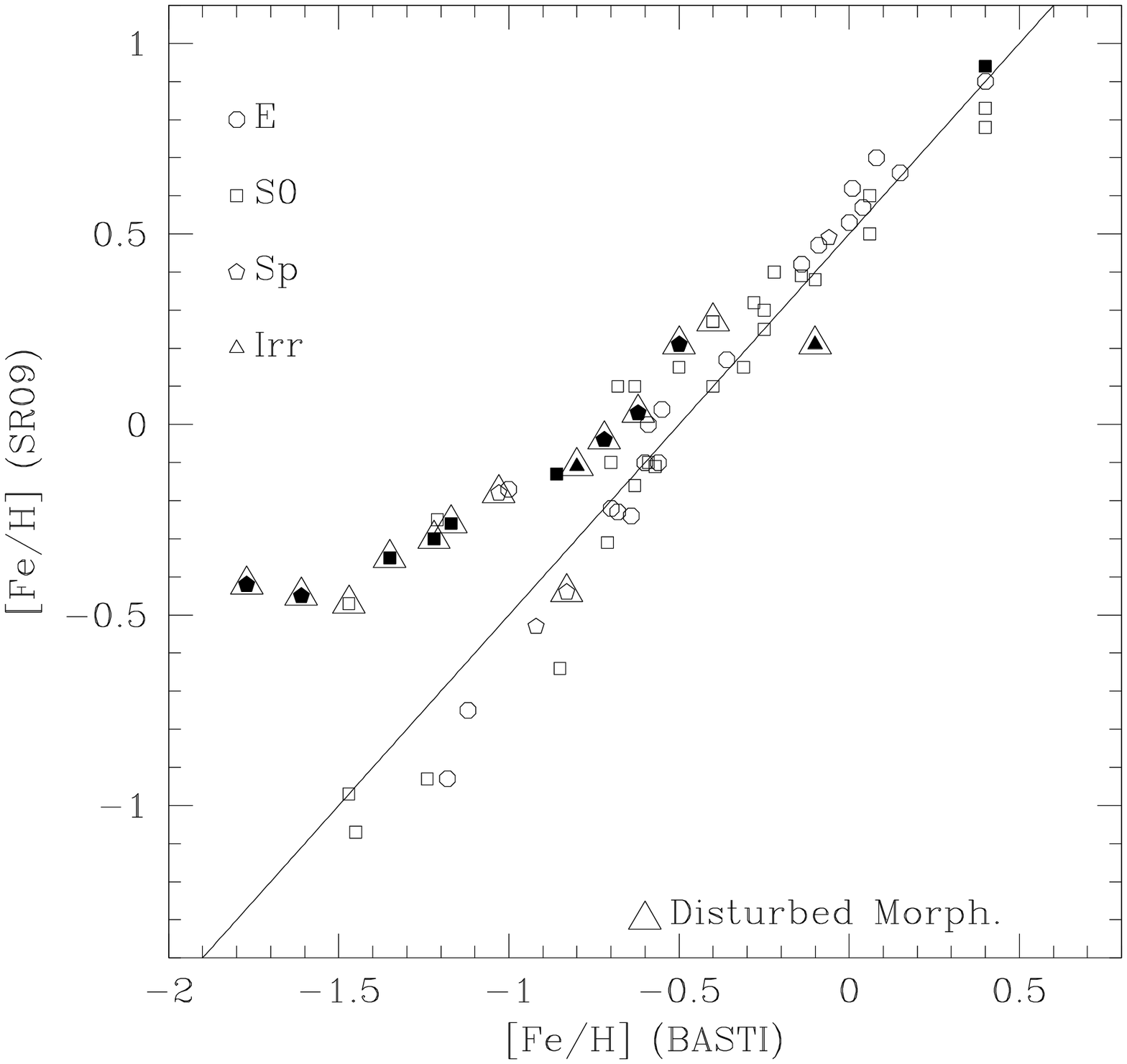}
\caption{Comparison between derived stellar population properties of the sample 
galaxies using respectively a grid of simple stellar population models and a grid 
of models based on the chemical evolution scenario of \citet{sr09a}. 
The upper panels show the optical/near-IR colour-colour diagrams for the sample 
galaxies compared  to the grid of models based on \citet{sr09a} chemical evolution 
scenario. Galaxies are separated by stellar masses (the left panel), with filled 
symbols showing galaxies with equivalent width of H$\alpha$ larger than 10\,\AA, 
and morphological type (the right panel). The lower panels show comparisons 
between the derived ages (left panel) and metallicities (lower panel) of the stellar
populations of the sample galaxies. Filled symbols show galaxies with equivalent 
width of H$\alpha$ larger than 10\,\AA. The solid line shows the relation
[Fe/H]$_{\rm SR09}$=[Fe/H]$_{\rm BASTI}$ + 0.5.    }
\label{bmk_jmk_comp}
\end{figure*}


\subsection{Single metallicity vs. multi-metallicity synthesis models}
\label{}

Galaxies are known to have experienced complex chemical evolution and star formation 
histories. To check if our choice of comparing galaxy colours to a grid of single age and
single metallicity stellar population models is biasing our results, we compare 
the optical/near-IR colours of the sample galaxies to a grid of models based on the 
chemical evolution scenario recently presented in \citet{sr09a}. 
This model assumes that galaxies are a sum of single stellar populations of metallicities 
following the infall scenario tuned to match the shape of the 
metallicity distribution functions of the Milky Way and NGC~5128. The metallicity 
distribution is allowed then to slide in the peak metallicity to produce a range of total 
metallicities per galaxy age. Each model produces a metallicity value based on the 
luminosity weighted average of each metallicity bin 
\cite[see][for a full description of the chemical evolution scenario]{sr09a}.

The upper panels of Fig.\,\ref{bmk_jmk_comp} show the comparison between a grid 
of stellar population synthesis models based on \citet{sr09a} chemical evolution 
scenario and the distribution of optical/near-IR colours of our sample galaxies. 
Galaxies are separated by stellar mass and equivalent width of H$\alpha$ emission 
(upper left panel), and morphological type (upper right panel). The overall features 
emerging from the comparison between the observed galaxy colours and the predicted 
grid are comparable to those noticed when the former were compared to the grid of 
simple stellar population models. The sequence of early-type galaxies, the population 
of intermediate age lenticular galaxies, and the population of (generally low mass) 
galaxies dominated by young stellar population are seen again. A careful comparison 
indicates, however, that differences exist. At a given age and metallicity, the 
multi-metallicity models tend to predict bluer colours than their equivalent BASTI models. 
This is due to the inclusion of a metal-poor tail that contribute more to colour than mean 
metallicity. The colours of the sample galaxies are then matched by models of higher 
metallicities and younger ages than when compared to predictions of single stellar 
population models. 

The lower panels of Fig.\,\ref{bmk_jmk_comp} show comparisons between the ages 
(lower left panel) and metallicities (lower right panel) of the sample galaxies derived 
by using the BASTI models and multi-metallicity models respectively. As suspected, 
the ages derived using the BASTI models are much older than those derived using 
the multi-metallicity models. When the latter models are considered, more than half 
of quiescent elliptical galaxies in the sample are predicted to be dominated by stars 
of young to intermediate ages, i.e., 2-5\,Gyr. This is surprising as the chemical evolution 
scenario of \citet{sr09a} was introduced to support the case that early-type galaxies on 
the red sequence are systematically old \cite[see][for a detailed discussion]{sr09a}.

Multi-metallicity stellar population models are found to over-estimate galaxy metallicities 
by a factor of about three compared to those derived using the BASTI models, over the 
entire metallicity range covered by passive galaxies in the sample. This is expected 
as the young ages found for elliptical galaxies using those models require solar to
super-solar metallicities to explain their colours. Differences in estimated metallicities 
for galaxies with ongoing star formation activity are even larger, up to about 1 dex 
at the metal-poor end of the metallicity range covered by the sample galaxies. 
This is due to the combined contribution of bright and blue young stars that dominated 
the light of those objects, and to the presence of blue stars populating the metal-poor 
tail of the metallicity distribution.

Despite those differences in the predicted ages and metallicities, the overall features 
of the distribution of the sample galaxies in the diagnostic diagrams discussed below 
are, however, reassuringly similar. As stellar population parameters derived by 
comparing galaxy colours to single stellar population synthesis models s have a 
straightforward interpretation, and due to the uncomfortably young ages of galaxies 
estimated when using the multi-metallicity models, we have decided to use the ages 
and metallicities derived using the BASTI models for the rest of the paper.

As galaxies contain mixed stellar populations in term of their ages and metallicities, 
the estimated stellar population properties by comparing their broad-band colours to 
those of synthetic single stellar populations will be luminosity-weighted mean values. 
Integrated galaxy stellar masses are estimated using the mass-to-light ratio vs. colour 
calibration of \citet{bj01}. These authors have argued that the stellar 
masses derived from their relation are robust against the effect of dust provided the 
dust vectors are parallel to the relation. The reason is that the under-estimate of stellar 
mass arising from the attenuation in luminosity will be compensated by the overproduction 
of stellar mass arising from the reddening in colour, therefore yielding comparable final 
stellar masses. This balance might apply for face-on systems, but could not be the case 
for highly inclined systems as the latter show a systematically higher ratio of 
attenuation to reddening. The overall consequence is that stellar masses could be 
under-estimated for a randomly oriented distribution by the use of the \citet{bj01}
colour vs. mass-to-light ratio if dust is not taken into account. \citet{driver07} have 
explored this in detail, and concluded that while the masses 
are modified somewhat the final stellar mass breakdown is not dramatically altered. 
The last three columns of Table~\ref{gal_prop_cal1} give the integrated stellar 
masses, the luminosity-weighted ages, and the luminosity-weighted metallicities 
for our sample of Abell 1367 galaxies.

\subsection{Stellar mass-metallicity relation}
\label{mass_met}

\begin{figure}
\includegraphics[clip=,width=0.5\textwidth]{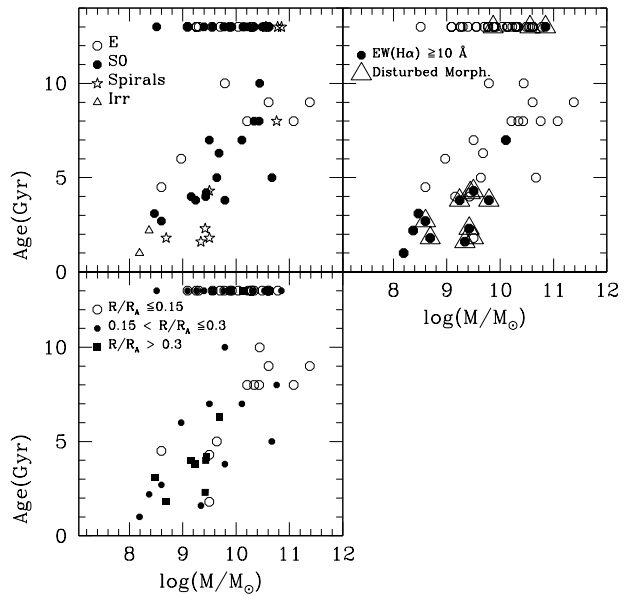}
\caption{The relationship between the luminosity-weighted stellar age and the stellar 
mass for our sample with galaxies separated by morphological type (upper left panel), 
H$\alpha$ emission line equivalent width and the presence of signs of disturbance 
(upper right panel), and radial distance (lower panel).}
\label{mst_age}
\end{figure}

Figure \ref{mst_age} shows the relationship between the luminosity-weighted age and the 
integrated stellar mass for our sample of Abell 1367 members. Galaxies are separated by 
morphological type in the upper left panel, equivalent width of H$\alpha$ emission line 
and the presence of signs of morphological disturbance in the upper right panel, and 
the cluster-centric distance normalized to the cluster Abell radius in the lower panel. 
As suspected from the distribution of galaxies in the optical/near-IR colour-colour 
diagram, elliptical galaxies show uniformly old ages across a factor of more than 
100 in integrated stellar mass. Out of the sub-sample of ellipticals, two objects stand 
out as their stellar contents appear to be dominated by populations of much younger 
luminosity-weighted ages, i.e., $\sim5$\,Gyr. The population of lenticular galaxies 
exhibits however a much clearer age vs. stellar mass relationship, extending from ages 
typical of old elliptical galaxies to young ages typical of those of late-type galaxies 
with young luminosity-weighted ages. Galaxies in the blue cloud, which are presently 
actively forming stars, show all young luminosity-weighted ages. The bulk of those 
galaxies are located at cluster-centric distances larger than 20 per cent of the Abell 
radius of the cluster.

\begin{figure*}
\includegraphics[clip=,width=0.45\textwidth]{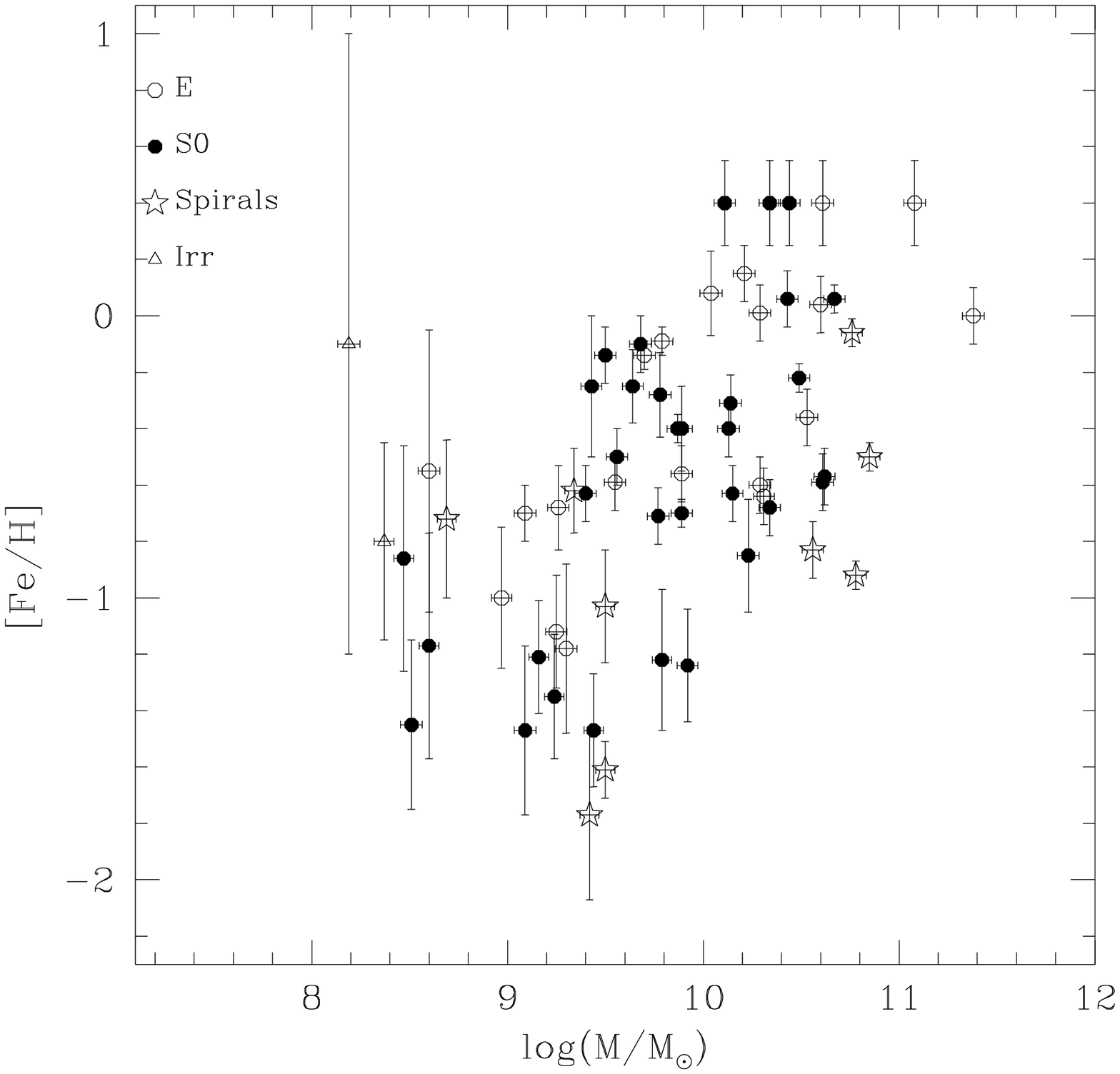}
\includegraphics[clip=,width=0.45\textwidth]{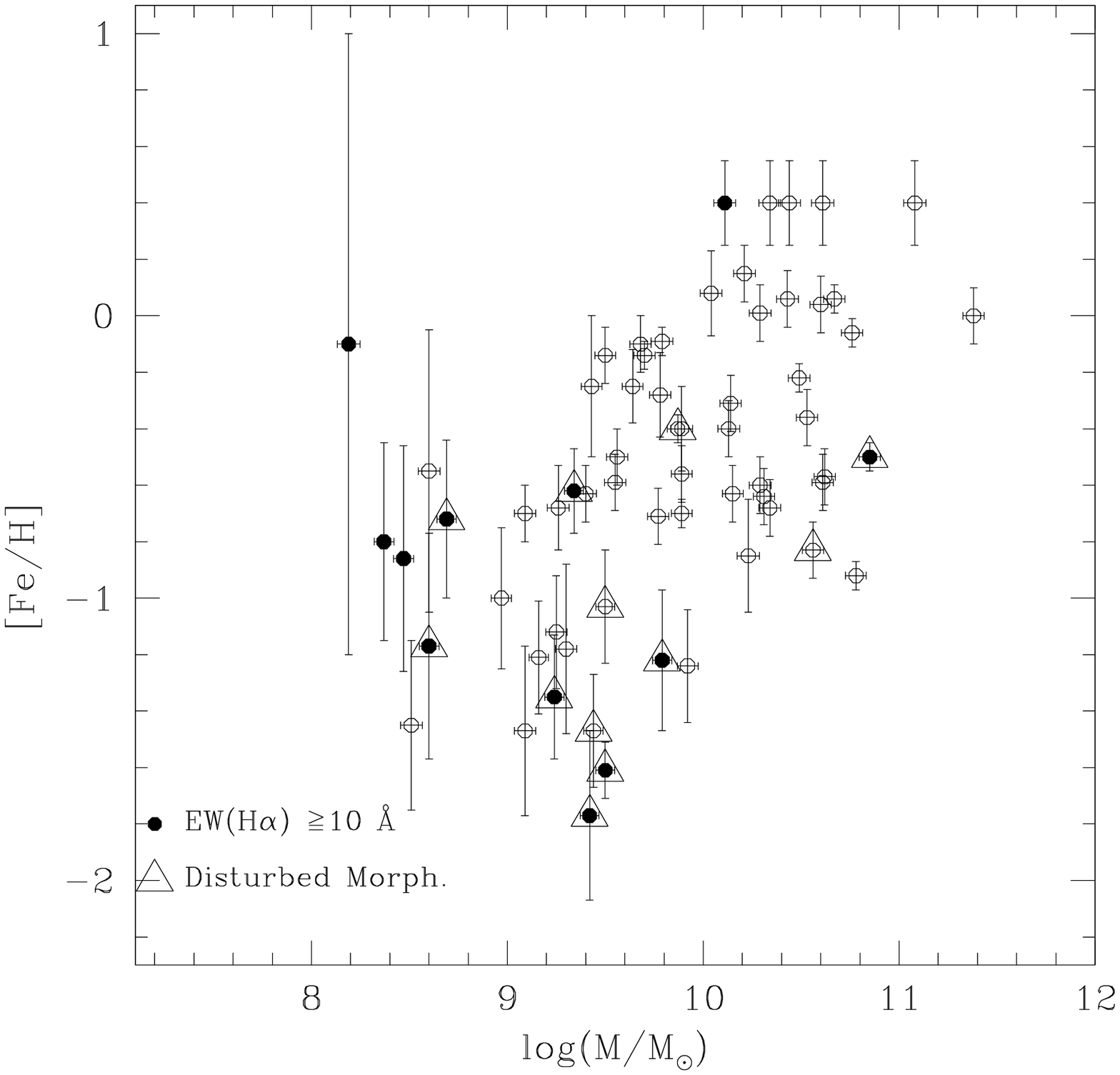}
\includegraphics[clip=,width=0.45\textwidth]{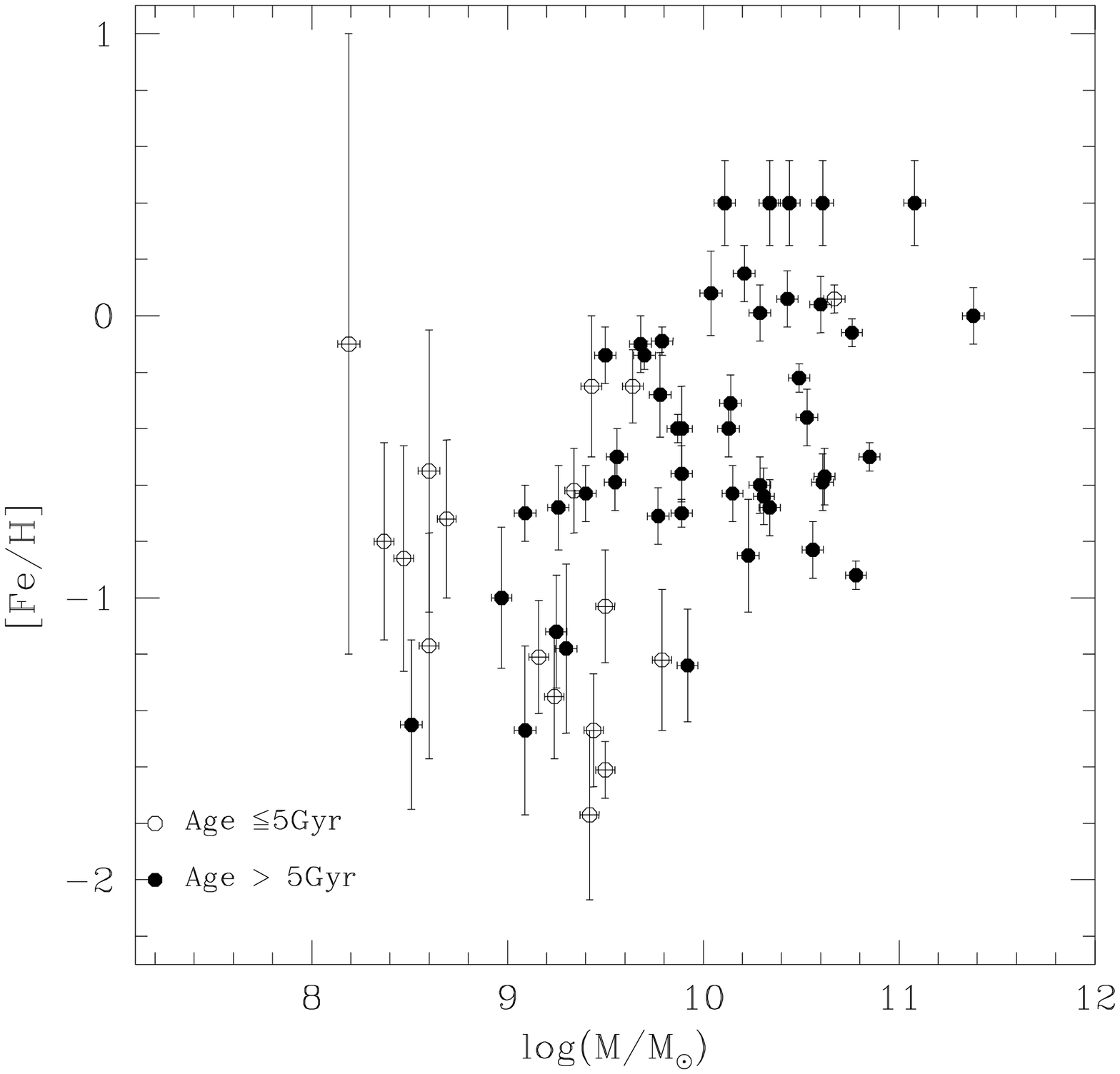}
\includegraphics[clip=,width=0.45\textwidth]{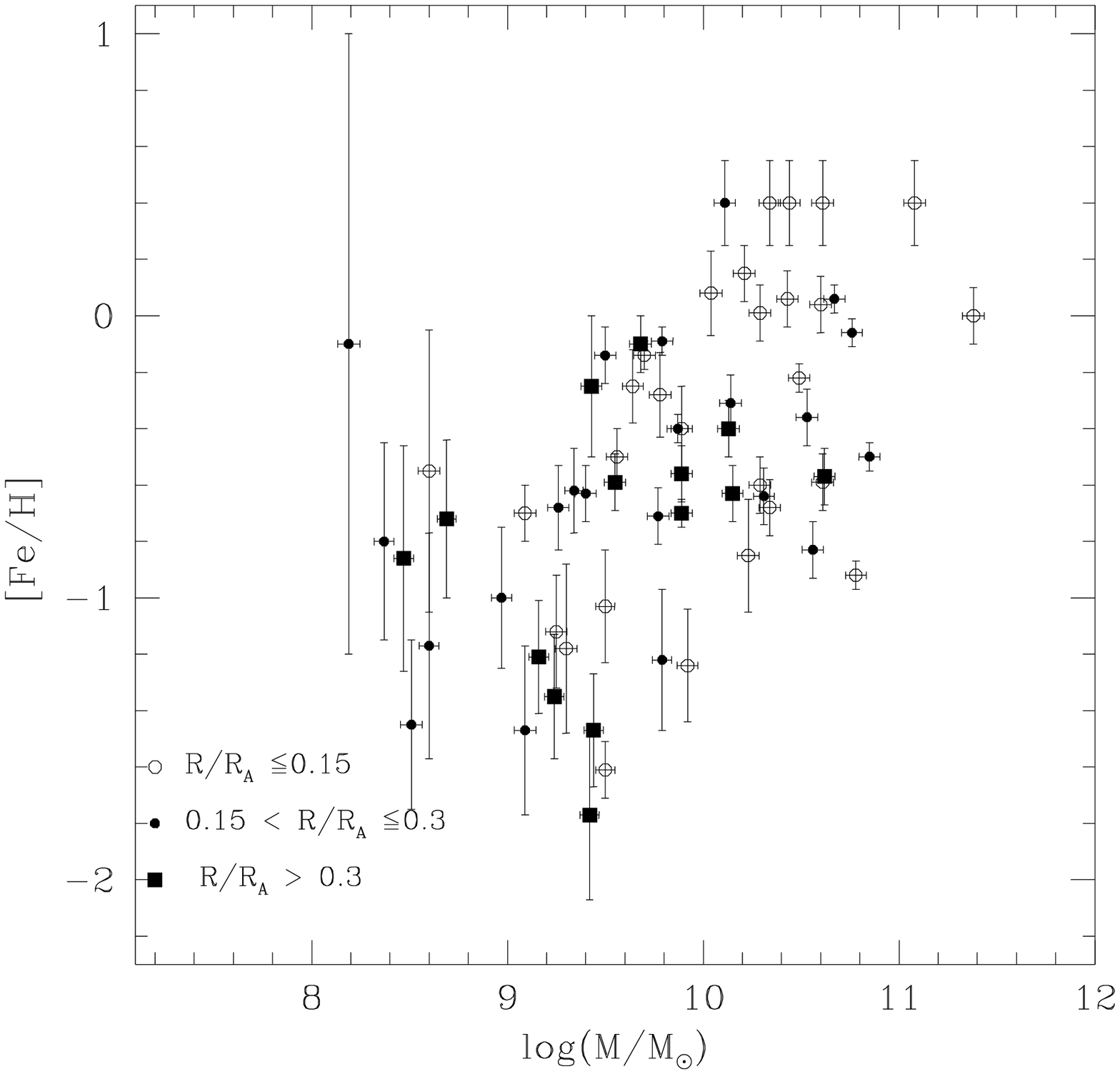}
\caption{The relationship between galaxy luminosity-weighted stellar metallicity and 
stellar mass for our sample with galaxies separated by morphological type (upper left 
panel), H$\alpha$ emission line equivalent width and the presence of signs of disturbance 
(upper right panel), luminosity-weighted stellar age (lower left panel), and 
cluster-centric distance (lower right panel).}
\label{mst_zset}
\end{figure*}

Fig.~\ref{mst_zset} shows the relationship between the luminosity-weighed metallicity 
and the integrated stellar mass for our sample of Abell 1367 members. Galaxies are 
separated by morphological type (upper left panel), equivalent width of H$\alpha$ 
emission line and the presence of signs of morphological disturbance (upper right panel), 
luminosity-weighted age (lower left panel), and the cluster-centric distance normalized 
to the Abell radius of the cluster (lower right panel). A clear trend between 
luminosity-weighted metallicity and stellar mass is apparent, in agreement with previous 
results for both field \cite[e.g.][]{gallazzi05} and cluster galaxies \cite[e.g][]{nelan05}.
The luminosity-weighted stellar metallicity of Abell 1367 
galaxies increases with stellar mass, from roughly 10 per cent of solar for galaxies 
with stellar masses below $10^{9}$~M$_{\odot}$ to about 2.5 times solar for those with 
stellar masses $\sim10^{11}$~M$_{\odot}$.  The scatter is clearly greater than the 
observational uncertainties, suggesting that the stellar mass is not the sole parameter 
that shapes the chemical histories of galaxies. For galaxies in the red sequence, neither 
age nor morphological type seem to be a factor in the relationship between integrated 
stellar mass and luminosity-weighted metallicity, as old elliptical galaxies cover 
a similar range in metallicity as lenticular galaxies which exhibit a spread in 
luminosity-weighted ages, and both populations have similar metallicity dispersions. 
Linear fits to red sequence elliptical and lenticular subsamples show that the slopes 
and the normalizations of the separate relationships are consistent with each other 
within the errors.

\begin{figure}
\includegraphics[clip=,width=0.5\textwidth]{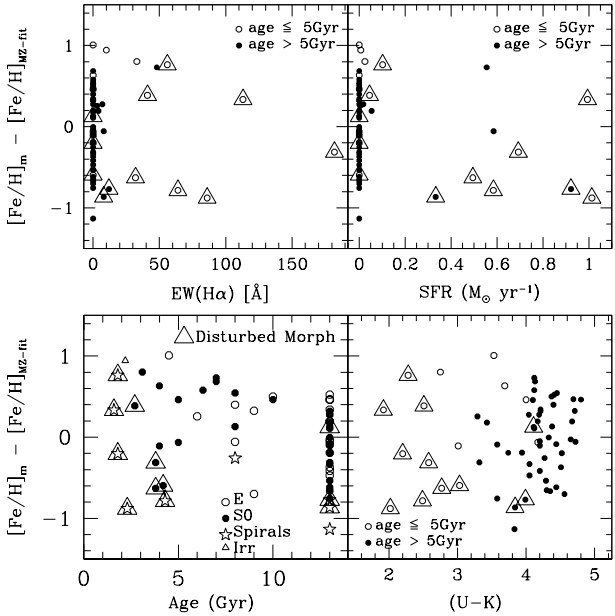}
\caption{Difference between the measured stellar metallicity and the stellar 
mass-metallicity relation as a function of the luminosity-weighted age (lower left panel), 
(U-K) colour (lower right panel), H$\alpha$ emission line equivalent width (upper left 
panel), and star formation rate (upper right panel), respectively.}
\label{resid_sfr_ew_cmr}
\end{figure}

Galaxies with ongoing star formation activity and/or with late-type morphological types 
show a different behaviour, however. They tend to have luminosity-weighted metallicities 
lower than those of the red sequence galaxies with similar integrated stellar masses. 
To investigate this further, Fig.\,\ref{resid_sfr_ew_cmr} shows the relationship between 
the residual from the linear least square bisector fit of the stellar mass-metallicity 
relation of the entire galaxy sample as a function of (U-K) colour (lower right panel), 
luminosity-weighted ages and morphological type (lower left panel), star formation rate 
(upper right panel), and H$\alpha$ emission line equivalent width (upper left panel). 
Galaxies with disturbed optical morphologies are marked by large open triangles in all 
panels. Apart from the lower left panel, galaxies with luminosity-weighted ages younger 
than 5\,Gyr are shown as open circles, older galaxies are displayed as filled circles. 
The star formation rates were estimated using the H$\alpha$ flux and the calibration 
of \citet{kennicutt98}. Fluxes were not corrected for either galaxy internal reddening nor 
from contamination from [NII]$\lambda6548,6583$ emission line. It can be clearly seen 
that the bulk of galaxies with disturbed morphologies and ongoing star formation, 
i.e., blue cloud galaxies, lie below the stellar mass-metallicity relation of the sample. 
No clear trend between the residuals from the stellar mass-metallicity and galaxy colour, 
mean age, emission line equivalent width, or star formation rate is apparent, suggesting 
none of them is driving alone the scatter around the stellar mass-metallicity relation.

The bulk of blue cloud galaxies in our sample exhibit disturbed morphologies. 
The observed tendency of blue galaxies in the cluster to be chemically underabundant 
relative to passive old galaxies in the red sequence could result from gas inflow induced 
by tidal interactions and/or minor mergers. The star formation activity associated with 
disturbed galaxies indicates that those galaxies still retain a substantial amount of 
gas which may be expected to be stripped by ram-pressure and/or harassment effect by 
extended exposure to the cluster environment. \citet{moss06} has found for an ensemble 
cluster formed from eight low-redshift clusters, that the majority of the infall galaxy 
population are in interacting or merging systems characterized by slow gravitational 
encounters. Merger numerical simulations predict gas inflows to occur in galaxy interactions
\cite[e.g.][]{mh96,barnes02}. Radial abundance gradients exist in most 
disk galaxies, such that gas phase metallicity decreases with increasing distance from the 
galactic centre \cite[e.g.][]{zaritsky94}. If gas inflows have triggered the star formation 
events as expected \cite[e.g.][]{bh96}, then those gas inflows may have carried 
less enriched gas from the outer regions of the galaxy, diluting the chemical abundances 
of the star-forming gas, thus lowering the metallicities of young stars. The recent numerical
simulations of equal-mass mergers of \citet{rupke10} confirm the hypothesis that 
merger-driven radial inflow of low metallicity gas from the outskirts of the interacting 
galaxies produce a dilution of nuclear abundances of interacting galaxies
\cite[see also][]{montuori10}. The recent observations of \citet{kewley10} indicate indeed
that the metallicity gradients for a sample of interacting galaxies are systematically 
shallower than for non-interacting galaxies. The interaction-induced gas inflow scenario 
was proposed to explain the offset of different classes of interacting/merging galaxies 
from the luminosity/mass-metallicity relationship of non-interacting galaxies. 
\citet{kewley06} and \citet{michel-dansac08} have shown that the interstellar 
gas of nearby optically-selected close galaxy pairs in the field is underabundant when 
compared to local star-forming galaxies of similar luminosities and stellar masses. 
The offset from the gas phase metallicity-luminosity for non-interacting galaxies 
appears to depend on the projected separation between the interacting galaxies and the 
burst strength \citep{kewley06}. \citet{peeples09} have identified a sample of metal-poor 
high-mass outliers from the local gas-phase mass-metallicity relation. 
These galaxies are found to exhibit disturbed morphologies, blue colours, and high star 
formation rates for their masses, implying that they are undergoing tidal interactions. 
Similarly, \citet{rupke08} have found that the interstellar gas of local luminous 
infrared galaxies is underabundant. The offset in abundance from the gas phase oxygen 
abundance vs. luminosity/mass relation is found to correlate with the total infrared 
luminosity, with a larger offset for ultra-luminous infrared galaxies than for luminous 
infrared galaxies. The offsets from the mass vs. gas phase metallicity relation in 
optically selected mergers is lower than for luminous and ultra-luminous infrared 
galaxies, suggesting that the offset from the properties of non-interacting galaxies 
depends on the strength and the stage of the interaction. The modest offset from the 
stellar mass-metallicity relation we observe for disturbed galaxies in the blue cloud 
sequence of Abell 1367 could reflect a continuation of the trends observed for 
equal-mass mergers to lower strength of galaxy interaction. Measuring gas phase 
metallicities of star-forming galaxies in our sample to compare with those of both 
morphologically undisturbed star-forming galaxies, and strongly disturbed galaxies 
would test this scenario.

\subsection{Radial dependence of stellar population properties}
\label{rad_dep}

For a cluster survey, the projected cluster-centric distance is a convenient parameter 
to describe the environment of a galaxy, although it may not be entirely free from 
introducing potential biases \cite[see][]{lane07}. In this section, we examine the 
dependence of stellar population properties of the cluster galaxies on the distance 
of the galaxy from the cluster centre, which has been taken to be at 
R.A.(J2000)=$11^{h} 44^{m} 29.5^{s}$ and dec(J2000)=$+19^{\circ} 50^{'} 21^{''}$. 
We express the cluster-centric radius in terms of the cluster Abell radius, taken 
to be $80^{'}$. 

\begin{figure*}
\includegraphics[clip=,width=0.33\textwidth]{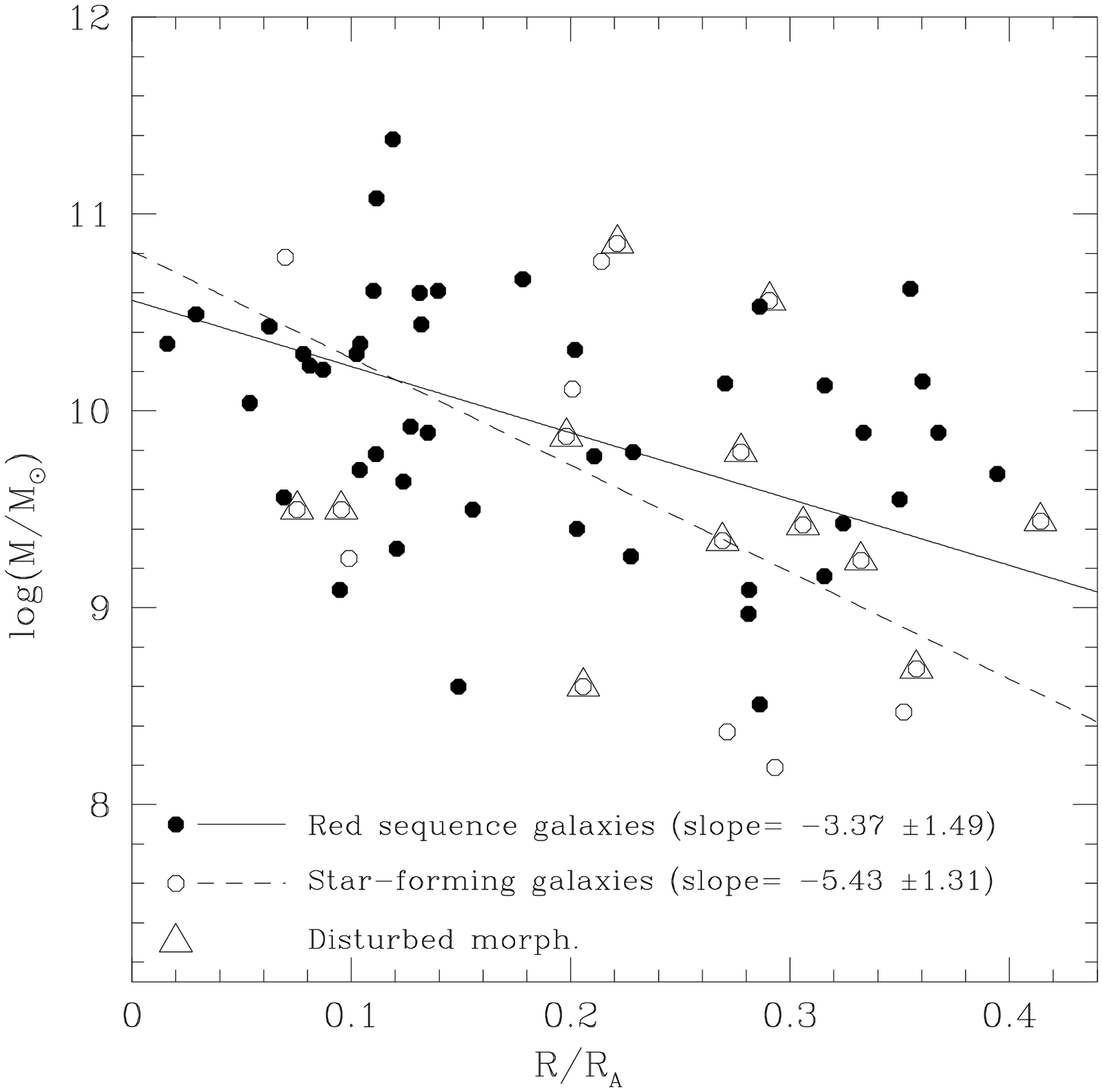}
\includegraphics[clip=,width=0.33\textwidth]{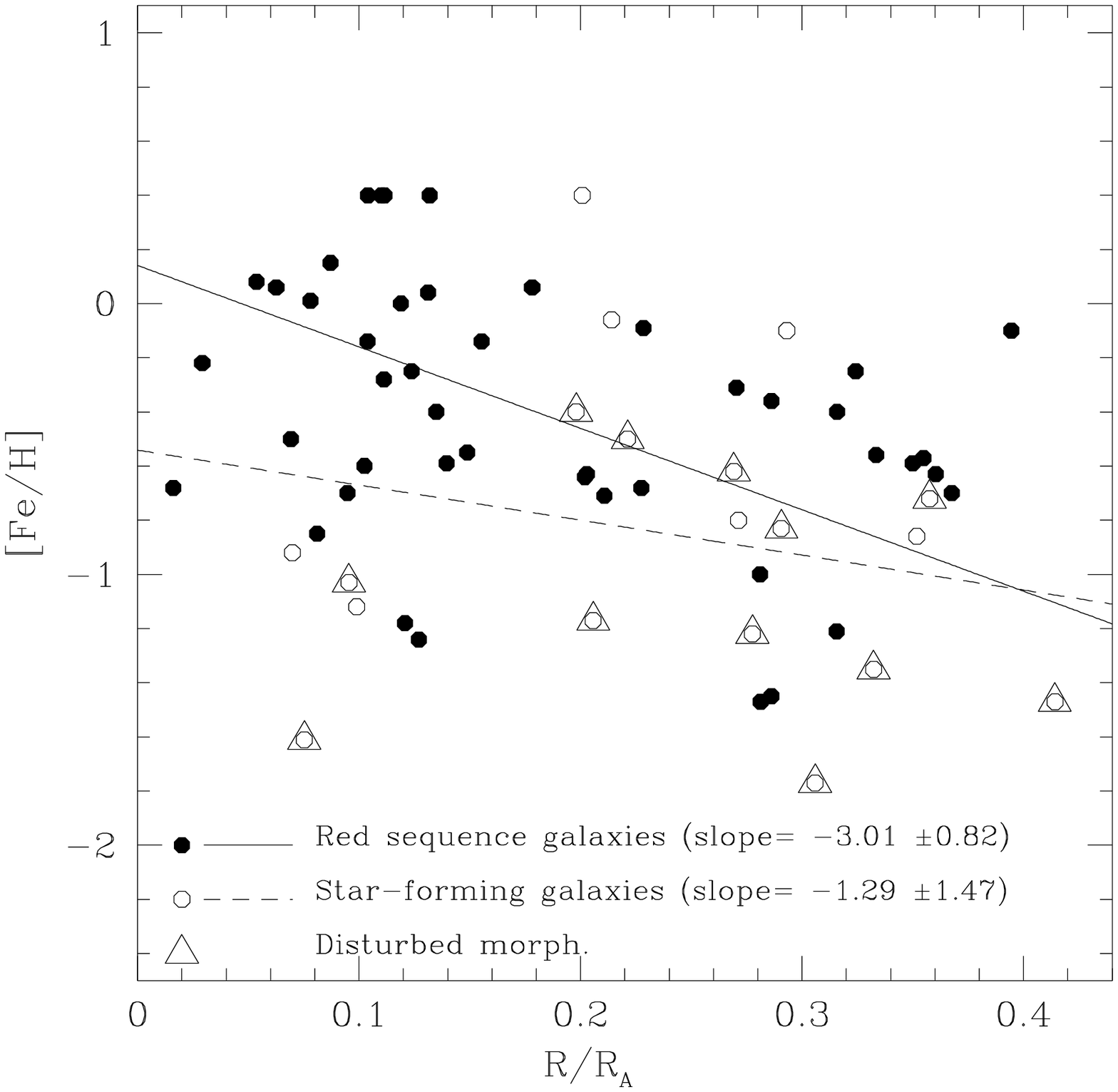}
\includegraphics[clip=,width=0.33\textwidth]{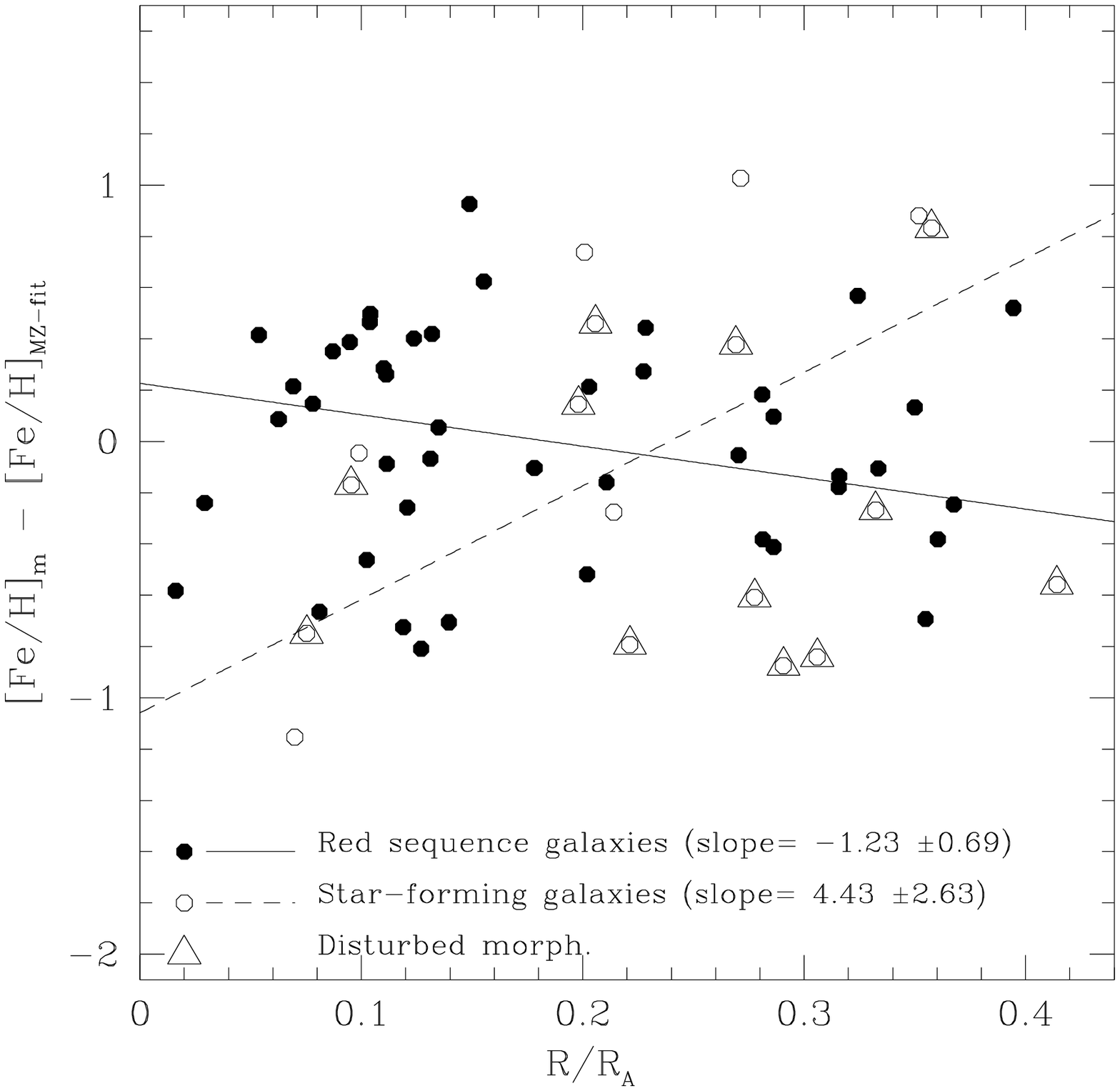}
\caption{Stellar mass, stellar metallicity, and residual from the stellar mass-metallicity 
relation as a function of the normalised projected radius for our sample of galaxies. 
In each panel, the solid line represents the least square bisector fit to the sub-sample 
of red sequence galaxies shown as filled circles, and the dashed line represents the fit 
to the sub-sample of blue cloud galaxies shown as open circles. Galaxies which display 
signs of morphological disturbances are flagged with large triangles.}
\label{resid_proj_RS_SF}
\end{figure*}

Figure \ref{resid_proj_RS_SF} shows the relationship between the cluster-centric radial distance, 
normalised to the Abell radius of the cluster, and the integrated stellar mass (left panel), the 
luminosity-weighted stellar metallicity (middle panel), and the residuals from the bisector fit 
of the stellar mass-metallicity relations (right panel). Galaxies in the sample are separated 
between those in the red sequence, shown as filled circles, and those in the blue cloud, shown as 
open circles. Galaxies that show signs of the presence of morphological disturbance are flagged 
with large open triangles. The residuals for each subsample are estimated using the linear 
least square bisector fit to the mass-metallicity relation of the subsample. In each panel, the 
solid line shows the best linear fit for red sequence galaxies, while the dashed line shows the 
fit to galaxies in the blue cloud. All galaxies are assigned equal weight in the fits, since the 
scatter is primarily intrinsic and the measurement errors similar for the bulk of the points. 
The slopes and the associated errors of the fit to the cluster-centric radial gradients are 
indicated for both subsamples.

As expected, more massive red sequence galaxies lie preferentially in the core of the cluster. 
As red sequence galaxies exhibit a clear mass-metallicity relation, a strong negative 
luminosity-weighted metallicity gradient as a function of the cluster-centric distance is present. 
The metallicity residuals for these galaxies do not show however a statistically significant 
correlation with the cluster-centric galaxy distance. Because the stellar populations of the 
cluster red sequence galaxies follow scaling relations with stellar mass, the residuals from 
these relations are comparing the inner and outer galaxy population at fixed stellar mass. 
On average, red sequence galaxies at 40 per cent of the Abell radius of the cluster are 
$\sim0.49\pm0.28$ dex poorer than galaxies of similar stellar masses in the cluster core. 
Similarly, blue cloud galaxies show a negative stellar mass gradient, however the 
luminosity-weighted metallicity shows no significant correlation with the cluster-centric radius. 
The residuals from the luminosity-weighted metallicity-mass relation of star-forming galaxies 
show no statistically significant correlation with the projected distance from the cluster centre. 
The lack of a negative correlation for the star forming galaxies could be caused by enhanced 
star formation, leading to lower luminosity-weighted metallicities (see section \ref{mass_met} 
above) in higher density regions of the cluster as compared with lower density regions
\cite[see][]{moss98,mw00}. Alternatively, given the small size of the our 
sample, projection effects could cause a dilution of the expected gradients, although such 
dilution effects are expected to be small in the sampled region within half the Abell radii 
\citep{moss98}.

Finally, neither red sequence galaxies nor star-forming galaxies show cluster-centric radial 
gradients for the residual from the relationship between luminosity-weighted age and stellar 
mass, over the radial range covered by the sample galaxies.

\begin{table*}
\caption[]{Properties of our sample of Abell 1367 galaxies}
\label{gal_prop_obs1}
\begin{tabular}{lllllllll}
\hline
SDSS-ID              & Type & Dis. &    (B-K)      &      (J-K)      & $\log$ (M/M$_{\odot}$) & Age (Gyr) & [Fe/H]  \\
\hline
J114229.18+200713.7   &  Irr     & n   &   2.08   $\pm$   0.12   &   0.42   $\pm$   0.15  &  &  &   \\      
J114239.31+195808.1   &  Sc     &y   &   2.61  $\pm$ 0.02   &   0.79  $\pm$ 0.02   &   8.69  $\pm$ 0.05   &   1.8  $\pm$ 0.3   &   -0.72  $\pm$ 0.28   \\
J114240.34+195717.0   &  S0     &   n   &   2.92  $\pm$ 0.03   &   0.78  $\pm$ 0.04   &   8.47  $\pm$ 0.05   &   3.1  $\pm$ 1.5   &   -0.86  $\pm$ 0.4   \\
J114246.11+195654.6   &  E       &   n   &   3.84  $\pm$ 0.01   &   0.86  $\pm$ 0.01   &   9.89  $\pm$ 0.05   &   13  $\pm$ 2   &   -0.56  $\pm$ 0.1   \\
J114251.93+195657.0   &  S0/a &   n   &   2.92  $\pm$ 0.01   &   0.76  $\pm$ 0.01   &   9.16  $\pm$ 0.05   &   4  $\pm$ 0.5   &   -1.21  $\pm$ 0.2   \\
J114253.19+201045.9   &  S0/a &   n   &   3.77  $\pm$ 0.01   &   0.9  $\pm$ 0.01   &   9.68  $\pm$ 0.05   &   6.3  $\pm$ 2   &   -0.10  $\pm$ 0.1   \\
J114256.45+195758.3   &  Scd   &   y   &   2.44  $\pm$ 0.01   &   0.71  $\pm$ 0.02   &   9.42  $\pm$ 0.05   &   2.3  $\pm$ 0.2   &   -1.77  $\pm$ 0.3   \\
J114258.69+195612.9   &  Irr     &   n   &   2.55  $\pm$ 0.07   &   0.85  $\pm$ 0.09   &   8.19  $\pm$ 0.06   &   1.0  $\pm$ 1.2   &   -0.10  $\pm$ 1.1   \\
J114301.18+195435.2   &  E      &   n   &   3.2  $\pm$ 0.02   &   0.78  $\pm$ 0.02   &   8.97  $\pm$ 0.05   &   6.0  $\pm$ 3   &   -1.00  $\pm$ 0.25   \\
J114306.32+195620.3   &  SB0  &   n   &   3.9  $\pm$ 0.01   &   0.88  $\pm$ 0.01   &   10.14  $\pm$ 0.05   &   13  $\pm$ 2   &   -0.31  $\pm$ 0.1   \\
J114313.09+200017.3   &  S,Pec   &   y   &   2.54  $\pm$ 0.01   &   0.8  $\pm$ 0.01   &   9.34  $\pm$ 0.05   &   1.6  $\pm$ 0.2   &   -0.62  $\pm$ 0.15   \\
J114313.27+200754.8   &  S0    &   n   &   3.47  $\pm$ 0.02   &   0.87  $\pm$ 0.03   &   9.43  $\pm$ 0.05   &   4.0  $\pm$ 1   &   -0.25  $\pm$ 0.25   \\
J114317.64+194657.6   &  E/S0 &   n   &   3.96  $\pm$ 0.01   &   0.92  $\pm$ 0.01   &   9.79  $\pm$ 0.05   &   10  $\pm$ 3   &   -0.09  $\pm$ 0.05   \\
J114321.95+195705.8   &  E      &   n   &   3.83  $\pm$ 0.01   &   0.84  $\pm$ 0.01   &   9.26  $\pm$ 0.05   &   13  $\pm$ 2   &   -0.68  $\pm$ 0.15   \\
J114324.55+194459.3   &  Sa    &   n   &   3.88  $\pm$ 0.01   &   0.92  $\pm$ 0.01   &   10.76  $\pm$ 0.05   &   8.0  $\pm$ 1   &   -0.06  $\pm$ 0.05   \\
J114324.88+203327.2   &  S0    &   n     &  &  &  & &      \\
J114336.23+195935.6   &  S0    &   n   &   3.81  $\pm$ 0.01   &   0.85  $\pm$ 0.01   &   9.4  $\pm$ 0.05   &   13  $\pm$ 2   &   -0.63  $\pm$ 0.1   \\
J114341.16+195311.1   &  S0    &   n   &   3.78  $\pm$ 0.01   &   0.9  $\pm$ 0.01   &   9.5  $\pm$ 0.05   &   7.0  $\pm$ 2   &   -0.14  $\pm$ 0.1   \\
J114341.54+200137.0   &  S0    &   y   &   2.68  $\pm$ 0.03   &   0.75  $\pm$ 0.03   &   8.6  $\pm$ 0.05   &   2.7  $\pm$ 0.5   &   -1.17  $\pm$ 0.4   \\
J114343.96+201621.7   &  S0/a &   n   &   3.91  $\pm$ 0.01   &   0.86  $\pm$ 0.01   &   10.62  $\pm$ 0.05   &   13  $\pm$ 2   &   -0.57  $\pm$ 0.1   \\
J114347.75+202148.0   &  S0    &   y   &   2.84  $\pm$ 0.01   &   0.72  $\pm$ 0.01   &   9.44  $\pm$ 0.05   &   4.2  $\pm$ 0.6   &   -1.47  $\pm$ 0.2   \\
J114348.89+201454.0   &  S0    &   y   &   2.82  $\pm$ 0.02   &   0.74  $\pm$ 0.02   &   9.24  $\pm$ 0.05   &   3.8  $\pm$ 0.9   &   -1.35  $\pm$ 0.22   \\
J114349.07+195806.4   &  Scd  &   y   &   3.46  $\pm$ 0.02   &   1.12  $\pm$ 0.02   &      &      &            \\
J114349.87+195834.8   &  S0    &   n   &   3.51  $\pm$ 0.01   &   1.06  $\pm$ 0.01   &      &      &             \\
J114350.12+195703.1   &  E      &   n   &   3.32  $\pm$ 0.05   &   0.82  $\pm$ 0.05   &   8.6  $\pm$ 0.06   &   4.5  $\pm$ 5   &   -0.55  $\pm$ 0.5   \\
J114351.33+195143.6   &  E/S0   &   n   &   3.44  $\pm$ 0.03   &   0.69  $\pm$ 0.04   &   9.3  $\pm$ 0.05   &   13  $\pm$ 2   &   -1.18  $\pm$ 0.3   \\
J114353.56+194421.7   &  S0   &   n   &   3.79  $\pm$ 0.01   &   0.88  $\pm$ 0.01   &   9.89  $\pm$ 0.05   &   13   $\pm$   3   &   -0.4  $\pm$ 0.15   \\
J114356.42+195340.4   &  E   &   n   &   4.21  $\pm$ 0.01   &   0.99  $\pm$ 0.01   &   11.08  $\pm$ 0.06   &   8.0  $\pm$ 3   &   0.40  $\pm$ 0.15   \\
J114357.44+195608.4   &  S0   &   n   &   3.66  $\pm$ 0.01   &   0.88  $\pm$ 0.01   &   9.64  $\pm$ 0.05   &   5.0  $\pm$ 2   &   -0.25  $\pm$ 0.13   \\
J114357.46+201802.0   &  S0   &   n   &   3.72  $\pm$ 0.01   &   0.85  $\pm$ 0.01   &   10.15  $\pm$ 0.05   &   13  $\pm$ 2   &   -0.63  $\pm$ 0.1   \\
J114357.50+195713.6   &  S0   &   n   &   4.26  $\pm$ 0.01   &   0.98  $\pm$ 0.01   &   10.44  $\pm$ 0.06   &   10  $\pm$ 3   &   0.40  $\pm$ 0.15   \\
J114357.70+201124.1   &  S0   &   n   &   3.19  $\pm$ 0.04   &   0.68  $\pm$ 0.04   &   9.09  $\pm$ 0.05   &   13  $\pm$ 2   &   -1.47  $\pm$ 0.3   \\
J114358.09+204822.9   &  Sa   &   n   &      &      &      &      &               \\
J114358.23+201107.9   &  S0   &   y   &   2.87  $\pm$ 0.02   &   0.75  $\pm$ 0.02   &   9.79  $\pm$ 0.05   &   3.8  $\pm$ 0.6   &   -1.22  $\pm$ 0.25   \\
J114358.83+195330.6   &  E   &   n   &   3.97  $\pm$ 0.01   &   0.91  $\pm$ 0.01   &   9.7  $\pm$ 0.05   &   13  $\pm$ 3   &   -0.14  $\pm$ 0.05   \\
J114358.95+200437.2   &  Sa   &   n   &   3.69  $\pm$ 0.01   &   1.08  $\pm$ 0.01   &      &         &        \\
J114359.57+194644.2   &  SB0   &   n   &   4.05  $\pm$ 0.01   &   0.97  $\pm$ 0.01   &   10.34  $\pm$ 0.05   &   8.0  $\pm$ 3   &   0.40  $\pm$ 0.15   \\
J114401.94+194703.9   &  Sc   &   y   &   2.49  $\pm$ 0.02   &   0.76  $\pm$ 0.02   &   9.5  $\pm$ 0.05   &   1.8  $\pm$ 0.2   &   -1.03  $\pm$ 0.2   \\
J114402.15+195659.3   &  E   &   n   &   4.02  $\pm$ 0.01   &   0.93  $\pm$ 0.01   &   11.38  $\pm$ 0.05   &   9.0  $\pm$ 3   &   0.00  $\pm$ 0.1   \\
J114402.15+195818.8   &  E   &   n   &   4.14  $\pm$ 0.01   &   0.94  $\pm$ 0.01   &   10.6  $\pm$ 0.05   &   13  $\pm$ 3   &   0.04  $\pm$ 0.1   \\
J114403.02+194424.8   &  S0   &   n   &   3.85  $\pm$ 0.01   &   0.89  $\pm$ 0.01   &   9.78  $\pm$ 0.05   &   13  $\pm$ 3   &   -0.28  $\pm$ 0.15   \\
J114403.21+194803.9   &  E   &   n   &   4.01  $\pm$ 0.01   &   0.95  $\pm$ 0.01   &   10.21  $\pm$ 0.05   &   8.0  $\pm$ 2   &   0.15  $\pm$ 0.1   \\
J114403.79+200556.1   &  S0   &   n   &   3.83  $\pm$ 0.01   &   0.83  $\pm$ 0.01   &   9.77  $\pm$ 0.05   &   13  $\pm$ 2   &   -0.71  $\pm$ 0.1   \\
J114405.46+195945.9   &  SB0   &   n   &   3.86  $\pm$ 0.01   &   0.86  $\pm$ 0.01   &   10.61  $\pm$ 0.05   &   13  $\pm$ 2   &   -0.59  $\pm$ 0.1   \\
J114405.74+201453.5   &  S0   &   n   &   4.02  $\pm$ 0.01   &   0.89  $\pm$ 0.01   &   10.13  $\pm$ 0.05   &   13  $\pm$ 2   &   -0.40  $\pm$ 0.1   \\
J114407.64+194415.1   &  E   &   n   &   3.84  $\pm$ 0.01   &   0.86  $\pm$ 0.01   &   10.29  $\pm$ 0.05   &   13  $\pm$ 2   &   -0.60  $\pm$ 0.1   \\
J114412.18+195633.9   &  E   &   n   &   3.77  $\pm$ 0.01   &   0.84  $\pm$ 0.02   &   9.09  $\pm$ 0.05   &   13  $\pm$ 2   &   -0.70  $\pm$ 0.1   \\
J114416.48+201300.7   &  E   &   n   &   4.09  $\pm$ 0.01   &   0.9  $\pm$ 0.01   &   10.53  $\pm$ 0.05   &   13  $\pm$ 2   &   -0.36  $\pm$ 0.1   \\
J114417.20+201323.9   &  Sa   &   y   &   3.7  $\pm$ 0.01   &   0.8  $\pm$ 0.01   &   10.56  $\pm$ 0.05   &   13  $\pm$ 2   &   -0.83  $\pm$ 0.1   \\
J114418.99+201812.9   &  E/S0   &   n   &   3.81  $\pm$ 0.01   &   0.86  $\pm$ 0.01   &   9.55  $\pm$ 0.05   &   13  $\pm$ 2   &   -0.59  $\pm$ 0.1   \\
J114420.42+195851.0   &  E/S0   &   n   &   4.25  $\pm$ 0.01   &   0.99  $\pm$ 0.01   &   10.61 0.06   &   9.0  $\pm$ 3   &   0.40  $\pm$ 0.15   \\
J114420.74+194933.3   &  SB0   &   n   &   4.03  $\pm$ 0.01   &   0.91  $\pm$ 0.01   &   10.49  $\pm$ 0.05   &   13  $\pm$ 2   &   -0.22  $\pm$ 0.05   \\
J114422.21+194628.2   &  E   &   n   &   4.19  $\pm$ 0.01   &   0.96  $\pm$ 0.01   &   10.04  $\pm$ 0.06   &   13  $\pm$ 3   &   0.08  $\pm$ 0.15   \\
J114425.12+194941.0   &  S0   &   n   &   3.87  $\pm$ 0.01   &   0.84  $\pm$ 0.01   &   10.34  $\pm$ 0.05   &   13  $\pm$ 2   &   -0.68  $\pm$ 0.1   \\
J114425.10+200628.1   &  E   &   n   &   3.73  $\pm$ 0.01   &   0.85  $\pm$ 0.01   &   10.31  $\pm$ 0.05   &   13  $\pm$ 2   &   -0.64  $\pm$ 0.1   \\
J114425.92+200609.6   &  S0   &   y   &   3.79  $\pm$ 0.01   &   0.88  $\pm$ 0.01   &   9.87  $\pm$ 0.05   &   13  $\pm$ 3   &   -0.4  $\pm$ 0.05   \\
J114428.36+194406.6   &  E   &   n   &   4.19  $\pm$ 0.01   &   0.95  $\pm$ 0.01   &   10.29  $\pm$ 0.06   &   13  $\pm$ 3   &   0.01  $\pm$ 0.1   \\
J114430.14+201944.5   &  S0   &   n   &   3.65  $\pm$ 0.01   &   0.83  $\pm$ 0.01   &   9.89  $\pm$ 0.05   &   13  $\pm$ 3   &   -0.70  $\pm$ 0.05   \\
J114430.55+200436.0   &  S0   &   n   &   3.84  $\pm$ 0.01   &   0.92  $\pm$ 0.01   &   10.67  $\pm$ 0.05   &   5.0  $\pm$ 0.8   &   0.06  $\pm$ 0.05   \\
J114432.12+200623.8   &  S0/a   &   n   &   3.99  $\pm$ 0.01   &   1  $\pm$ 0.01   &   10.11  $\pm$ 0.05   &   7.0  $\pm$ 5   &   0.40  $\pm$ 0.15   \\
J114436.55+194505.7   &  S0    & n   &   3.74  $\pm$ 0.01   &   0.86  $\pm$ 0.01   &   9.56  $\pm$ 0.05   &   13  $\pm$ 3   &   -0.50  $\pm$ 0.1   \\
J114446.62+194528.3   &  S0   &   n   &   3.99  $\pm$ 0.01   &   0.77  $\pm$ 0.01   &   10.23  $\pm$ 0.06   &   13  $\pm$ 2   &   -0.85  $\pm$ 0.2   \\
\hline
\end{tabular}
\end{table*}

\begin{table*}\addtocounter{table}{-1}
\caption[]{Properties of our sample of Abell 1367 galaxies (continued).}
\label{gal_prop_cal1}
\begin{tabular}{lllllllll}
\hline
SDSS-ID              & Type & Dis. &    (B-K)      &      (J-K)      & $\log$ (M/M$_{\odot}$) & [Fe/H] & Age (Gyr)  \\
\hline
J114447.03+200730.3   &  Sbc   &   y   &   3.79  $\pm$ 0.01   &   0.87  $\pm$ 0.01   &   10.85  $\pm$ 0.05   &   13  $\pm$ 3   &   -0.5  $\pm$ 0.05   \\
J114447.28+201247.4   &  S0   &   n   &   3.29  $\pm$ 0.03   &   0.66  $\pm$ 0.04   &   8.51  $\pm$ 0.06   &   13  $\pm$ 2   &   -1.45  $\pm$ 0.3   \\
J114447.44+195234.9   &  S0   &   n   &   3.99  $\pm$ 0.01   &   0.94  $\pm$ 0.01   &   10.43  $\pm$ 0.05   &   8.0  $\pm$ 1.5   &   +0.06  $\pm$ 0.1   \\
J114447.79+194624.3   &  Sc   &   y   &   2.81  $\pm$ 0.01   &   0.72  $\pm$ 0.01   &   9.5  $\pm$ 0.05   &   4.3  $\pm$ 0.5   &   -1.61  $\pm$ 0.1   \\
J114447.95+194118.6   &  SB0/a  &   n   &   3.33  $\pm$ 0.01   &   0.69  $\pm$ 0.01   &   9.92  $\pm$ 0.05   &   13  $\pm$ 2   &   -1.24  $\pm$ 0.2   \\
J114449.16+194742.2   &  Sa   &   n   &   3.64  $\pm$ 0.01   &   0.78  $\pm$ 0.01   &   10.78  $\pm$ 0.05   &   13  $\pm$ 2   &   -0.92  $\pm$ 0.05   \\
J114449.62+195627.9   &  E   &   n   &   3.4  $\pm$ 0.01   &   0.73  $\pm$ 0.02   &   9.25  $\pm$ 0.05   &   13  $\pm$ 2   &   -1.12   $\pm$   0.2   \\
J114450.31+195903.7   &  SB0/a   &   n   &   2.65  $\pm$ 0.08   &   0.87  $\pm$ 0.1   &      &      &         \\
J114451.09+194718.2   &  Irr   &   n   &   1.6  $\pm$ 0.21   &   0.58  $\pm$ 0.26   &      &      &         \\
J114452.40+201117.0   &  Irr   &   n   &   2.68  $\pm$ 0.03   &   0.78  $\pm$ 0.03   &   8.37  $\pm$ 0.05   &   2.2  $\pm$ 0.5   &   -0.8  $\pm$ 0.35   \\
\hline
\end{tabular}
\end{table*}

\begin{figure}
\includegraphics[clip=,width=0.5\textwidth]
{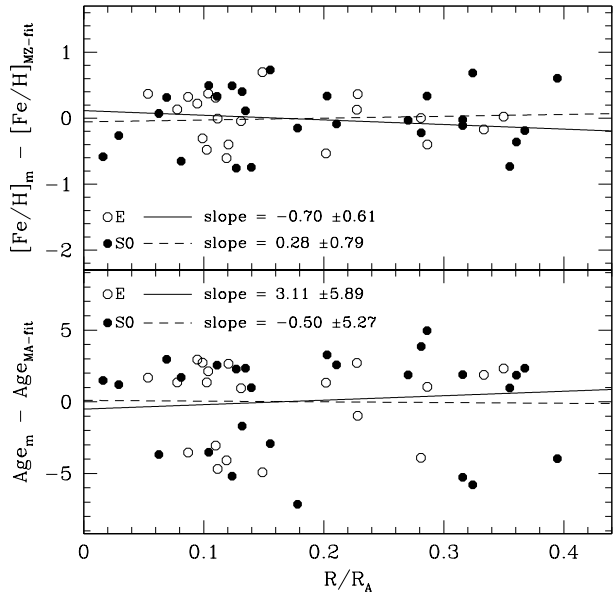}
\caption{Residuals from the stellar mass-metallicity relation as a 
function of the normalised projected radius for red sequence ellipticals 
and S0s respectively.}
\label{res_feh_age_ellip_S0}
\end{figure}

Ellipticals and lenticulars populating the red sequence do appear to be dominated by 
different stellar populations (see Fig.~\ref{bmk_jmk} and \ref{mst_age}), suggesting that 
they might have experienced different assembly histories. It is legitimate then to check 
how their stellar population properties taken separately change radially within the cluster. 
Figure \ref{res_feh_age_ellip_S0} shows the relationships between the cluster-centric 
distance normalized to the Abell radius of the cluster and the residuals from the stellar 
mass-metallicity relations (the upper panel) and the stellar mass-age relations (the lower 
panel) for ellipticals and lenticulars in the red sequence shown as open and filled circles 
respectively. The fits to the stellar mass vs. metallicity/age relations are done separately 
for elliptical and lenticular subsamples. The slopes and their associated errors of the 
best linear fits to the relations between the residuals and the normalized cluster-centric 
radial distance are indicated. No statistically significant trends are found between the 
residuals from the mass-metallicity/mass-age relations and the cluster-centric galaxy 
distance for neither red sequence ellipticals nor lenticulars. This indicates that no 
significant differences in the mean luminosity-weighted metallicities and ages are present, 
at a fixed stellar mass, between the red sequence galaxy sub-populations distributed 
within the 40 per cent of the Abell radius of the cluster covered by our survey. 
The lack of age differences between red sequence galaxies at different locations in 
the cluster could be a genuine property of the red galaxy population of the cluster, 
but it could alternatively be caused by the low sensitivity of broad-band colours to small 
age variation for old stellar populations. Extending the spatial coverage of the cluster 
is needed however to investigate firmly the differences between galaxy sub-populations 
of the red sequence.

\begin{figure}
\includegraphics[clip=,width=0.5\textwidth]{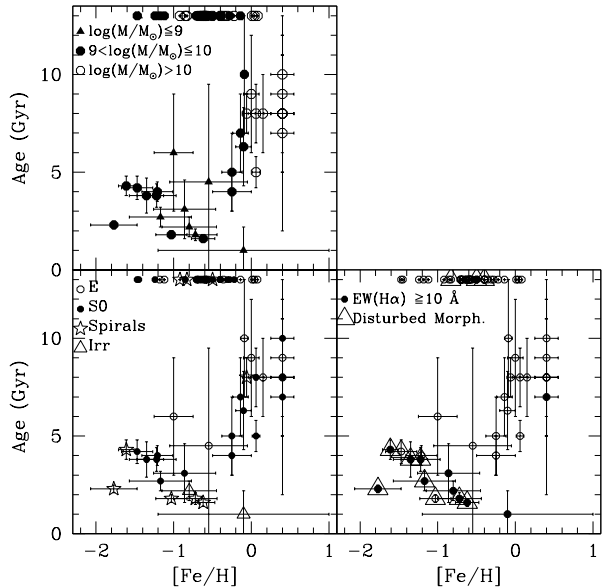}
\caption{The relationship between the luminosity-weighted stellar metallicity and age 
for our sample with galaxies separated by morphological type (lower left panel), 
H$\alpha$ emission line equivalent width and the presence of signs of morphological 
disturbance (lower right panel), and stellar mass (upper panel).}
\label{age_feh_morph_mass_ew}
\end{figure}

Figure \ref{age_feh_morph_mass_ew} shows the relationship between the luminosity-weighted 
ages and metallicities for the galaxies in our sample, separated by morphological type 
shown in the lower left panel, equivalent width of H$\alpha$ emission line shown in the 
lower right panel, and integrated stellar mass shown in the upper panel. Cluster galaxies 
divide clearly into multiple populations by metallicity and age. The first population is 
the classic old red sequence galaxies, with ages older than 10-13\,Gyr, with predominantly 
early-type morphologies, and mean metallicities covering a wide range from $\sim10$ per 
cent solar to slightly super-solar. The second group of cluster galaxies, dominated by red 
sequence lenticular galaxies, covers a similar metallicity range to the first group, but 
is younger with luminosity-weighted ages ranging between 5 and 10\,Gyr. The third group 
of cluster members contains the younger galaxies populating the blue cloud in the 
colour-magnitude diagrams, with luminosity-weighted metallicities extending to lower 
values than those typical for galaxies in the first two groups. Although galaxy 
luminosity-weighted metallicity is correlated with galaxy mass, it is uncorrelated with 
luminosity-weighted age for the three populations. \citet{rakos08} have used narrow-band 
colours combined with principal component analysis \citep{rs05} to estimate 
the luminosity-weighted stellar metallicities and ages for galaxies in a sample of nearby 
galaxy clusters. They have found cluster early-type galaxies to be divided into two distinct 
populations, an old galaxy population dominated by primordial stellar populations and a 
second younger galaxy population, well separated by a gap in age of about 2\,Gyr from the 
oldest galaxies, in agreement with our findings. The third population of cluster galaxies 
we have identified here is not detected in their analysis, as their technique fails to 
determine the luminosity-weighted properties for stellar populations dominated by stars 
younger than $\sim3$\,Gyr. In addition, galaxies with ongoing star formation or those 
suspected to have large fractions of young stars identified by their colours and/or 
strong emission lines were not included in their analysis \cite[see also][]{rakos07}. 
In their sample of nearby galaxy clusters, the populations of bright blue galaxies and 
faint starburst galaxies are not as populated as in the case of Abell 1367 
\cite[see][]{odell02,rakos96}.

\section{Discussion}

A number of studies have analysed absorption line-strength indices of cluster galaxies, 
and claimed the detection of variation of galaxy properties as a function of cluster-centric 
radius. \citet{guzman92} reported a positive offset of the zero-point of the relationship 
between Mg$_2$ index and galaxy velocity dispersion for massive elliptical galaxies between 
the core and the outskirts of the Coma cluster. At a fixed velocity dispersion, the Mg$_2$ 
line strength is weaker at larger cluster-centric radius. \citet{carter02} have used a sample 
of Coma cluster galaxies extending out to the virial radius, without morphological or colour 
selection and distributed over a luminosity range similar to galaxies in our sample, to 
report a significant radial decrease in the Mg$_2$ index, and an increase of the H$\beta$ 
index. Carter et al. have attributed the observed radial trends to change in metal abundance 
at fixed stellar mass between the core and the outskirts of the cluster. 

\citet{smith08}, in a study of 75 red-sequence dwarf galaxies in the Coma cluster, find 
that the variations in line strength indices are driven by variations in age and in iron 
abundance, with $\alpha$-element abundances being independent of radius. Dwarf galaxies 
in the outer regions of the Coma cluster are on average younger and more iron-enriched than 
those in the core, at a given luminosity. These results must be treated with caution however, 
as both of the Coma studies discussed above observed the same, possibly atypical, outer region 
of the cluster to the southern-west of the core.

Using a sample of $\sim3000$ red sequence galaxies dominated by $L_{*}$ galaxies in $\sim90$ 
nearby clusters, \citet{smith06} found strong evidence for stronger Balmer lines and 
weaker light-element features, at fixed mass, but no radial dependence for iron-dominated 
indices over a range from the cluster centre to the virial radius. The radial gradients 
detected for the synthetic cluster formed by stacking galaxies in all clusters in their 
sample, are significantly shallower than those found for the Coma cluster, with different 
relative patterns for different spectral indices. \citet{smith06} have argued that the 
observed cluster-centric radial gradients are better explained by a change in the mean age 
of the dominant stellar populations rather than in metallicity between galaxies in the 
cores and the outskirts of clusters. Red sequence galaxies at the virial radius have on 
average younger ages than galaxies of the same velocity dispersion situated near the cluster 
centres. More recently, \citet{rakos07} estimated the cluster-centric radial metallicity 
gradients for a sample of passive, red galaxies in Abell 1185 cluster. They have found a 
significant cluster-centric metallicity gradient, with mean metallicities decreasing by 
$\sim0.2$ dex within 40 per cent of the Abell radius of the cluster.

Our analysis shows that the average luminosity-weighted properties of Abell 1367 red 
sequence galaxies do not show a significant change, at fixed stellar mass, with the location 
within the cluster over a radial distance ranging from the cluster centre to a radius of 
approximately half the cluster Abell radius. The complex structure and dynamics of Abell 1367 
could lead however to concerns about the extent to which the region covered by our survey 
is representative of the entire galaxy population of the cluster. 
In contrast with \citet{smith06}, the red sequence ellipticals in the core of Abell 1367 
appear to be of similar average luminosity-weighted ages than those in the outer regions of 
the cluster. To account for the observed gradients in star formation rate measured for the 
CNOC cluster sample, \citet{balogh00} have proposed a model in which star-forming galaxies 
fall into rich clusters and their star formation is gradually quenched by ram pressure and 
tidal stripping, removing their gas over a few Gyr. This mechanism would be expected to lead 
to a radial age gradient. In the semi-analytical models of \citet{delucia06}, the mean 
luminosity-weighted age of bulge-dominated galaxies falls from $\sim12$\,Gyr at the cluster 
centres to $\sim10.5$\,Gyr at the virial radius. Some of this effect is due to mass 
segregation however, as more massive galaxies are preferentially located at the core of 
the cluster. The restricted range of cluster-centric radial distance covered by our survey, 
i.e., out to approximately 40 per cent of the virial radius of the cluster, combined with 
the low sensitivity of broad-band colours to age variation for old stellar populations, 
projection effects, and the size of the sample means we cannot rule out the presence of a 
radial age gradient. A comprehensive answer to these concerns should result from an extension 
of our survey out to the cluster infall region to test if galaxies in other parts of Abell 1367 
display similar radial trends.

\section{Summary}
\label{summary}

Using deep optical/near-IR broad-band and narrow-band imaging data, we investigate the 
stellar population properties for a sample of Abell 1367 galaxies with spectroscopically 
determined memberships and visual morphological classifications. Our survey samples 
galaxies with integrated stellar masses ranging from a few $10^{8}$\,M$_{\odot}$ up to a 
few $10^{11}$\,M$_{\odot}$. We compare the optical/infrared colours of galaxies in our sample 
with the predictions of simple stellar population models, spanning a range in metallicity 
and age, to determine the luminosity-weighted parameters of their stellar contents. 
The colours of galaxies morphologically classified as ellipticals are well described 
by a sequence of varying metallicity for the stellar population at a constant old age. 
In contrast, lenticulars along the red sequence show a large spread in their optical/near-IR 
colours, indicating a spread in the ages and the metallicities of their dominant stellar 
components. 

The stellar population parameters of the cluster galaxies show systematic trends as a 
function of galaxy stellar mass. We find that the luminosity-weighted properties of galaxy 
stellar contents correlate with stellar mass in the sense that less massive galaxies are 
dominated by stellar populations of younger ages and lower stellar metallicities than their 
massive counterparts. There is however a segregation in the stellar metallicity between 
quiescent and star forming galaxies. Most of the galaxies which show signs morphological 
disturbance and have ongoing star formation activity, lie systematically below the stellar 
mass-metallicity relation defined by galaxies populating the red sequence. 
The low luminosity-weigthed stellar metallicities of disturbed galaxies could be attributed 
to tidally-driven gas inflows from the galaxy outskirts which dilute the star-forming gas, 
leading to less chemically evolved young stars.

We have investigated the cluster-centric radial dependence of the properties of the stellar 
contents of galaxies. After allowing for the mass dependence of the stellar population 
properties, we find that the luminosity-weighted properties are not correlated to the 
cluster-centric radius for galaxies in both the red sequence and the blue cloud, over 
the radial distance covered by the sample galaxies, i.e., from the cluster centre to 
approximately half the cluster Abell radius. 

The cluster red sequence galaxies are found to be divided into two distinct populations. 
The first is a subpopulation of elliptical galaxies with uniformly old stellar populations 
with average ages similar to the age of the Universe. The second sub-population is 
comprised of red sequence lenticulars with younger ages, for which an age-mass relation 
is present. None of these two galaxy sub-populations of the red sequence show neither 
metallicity nor age cluster-centric gradients, over the radial distance covered by the 
survey. The third cluster subpopulation is that of galaxies where star formation is still 
ongoing and which are morphologically disturbed.

\bsp

\label{lastpage}


\begin{thebibliography}{}
\bibitem[Balogh et al. (2000)]{balogh00} 
Balogh M. L., Navarro J. F., Morris S. L., 2000, ApJ, 540, 113
\bibitem[Barnes \& Hernquist (1996)]{bh96} 
Barnes J. E., \& Hernquist L., 1996, ApJ, 471, 115
\bibitem[Barnes (2002)]{barnes02} 
Barnes, J. E. 2002, MNRAS, 333, 481
\bibitem[Bell \& de Jong (2001)]{bj01} 
Bell E. F., de Jong R. S., 2001, ApJ, 550, 212
\bibitem[Boselli et al. (1994)]{boselli94} 
Boselli A., Gavazzi G., Combes F., Lequeux J., Casoli F., 1994, A\&A, 285, 69
\bibitem[Boselli \& Gavazzi (2006)]{bg06} 
Boselli A., Gavazzi G., 2006, PASP, 118, 517
\bibitem[Bressan et al. (1994)]{bressan94} 
Bressan A., Chiosi C., Fagotto F., 1994, ApJS, 94, 63
\bibitem[Burstein et al. (1984)]{burstein84} 
Burstein D., Faber, S. M., Gaskell C. M., Krumm N., 1984, ApJ, 287, 586
\bibitem[Buta (1996)]{buta96} 
Buta R., 1996, AJ, 111, 591
\bibitem[Butcher \& Oemler (1984)]{bo84} 
Butcher H., Oemler A. Jr.,  1984, ApJ, 285, 426
\bibitem[Carter et al. (2002)]{carter02} 
Carter D., Mobasher B., Bridges T. J., Poggianti B. M., 
Komiyama Y., Kashikawa N., Doi M., Iye M., Okamura S., Sekiguchi M., 
Shimasaku K., Yaki M., Yasuda N. 2002, ApJ, 567, 772
\bibitem[Cortese et al. (2004)]{cortese04} 
Cortese L., Gavazzi G., Boselli A., Iglesias-Paramo J., Carrasco L., 2004, A\&A, 425, 429
\bibitem[De Lucia et al. (2006)]{delucia06} 
De Lucia G., Springel V., White S. D. M., Croton D., Kauffman G., 2006, MNRAS, 366, 499
\bibitem[de Vaucouleurs \& Longo (1998)]{vl98} 
de Vaucouleurs A., Longo G. 1988, Catalogue of Visual and Infrared Photometry of Galaxies 
from 0.5 micrometer to 10 micrometer (1961--1985).  Univ. of Texas, Austin
\bibitem[Dickey \& Gavazzi (1991)]{dg91} 
Dickey J. M., Gavazzi G., 1991, ApJ, 373, 347
\bibitem[Donnelly et al. (1998)]{donnelly98} 
Donnelly R. H., Markevitch M., Forman W., Jones C., David L. P.,  Churazov E., Gilfanov M., 
1998, ApJ, 500, 138
\bibitem[Driver et al. (2007)]{driver07} 
Driver S. P., Popescu C. C., Tuffs R. J., Liske J., Graham A. W., Allen P. D., de Propris R., 
2007, MNRAS, 379, 1022
\bibitem[Dressler (1980)]{dressler80} 
Dressler A., 1980, ApJ, 236, 772
\bibitem[Dressler et al. (1997)]{dressler97} 
Dressler A., et al., 1997, ApJ, 490, 577
\bibitem[Faber (1973)]{faber73} 
Faber, S. M. 1973, ApJ, 179, 731
\bibitem[Gallazzi et al. (2005)]{gallazzi05} 
Gallazzi A., Charlot S., Brinchmann J., White S. D. M., Tremonti C. A., 2005, MNRAS, 362, 41 
\bibitem[Gao et al. (2004)]{gao04} 
Gao L., White S. D. M., Jenkins A., Stoehr F., Springel V., 2004, MNRAS, 355, 819
\bibitem[Gavazzi et al. (2001)]{gavazzi01} 
Gavazzi G., Boselli A., Mayer L., Iglesias-Paramo J., V\'ilchez J. M., Carrasco L., 2001, 
ApJ, 563L, 23G
\bibitem[Gavazzi et al. (2003)]{gavazzi03} 
Gavazzi G., Cortese L., Boselli A., Iglesias-Paramo J., V\'ilchez J. M., Carrasco L., 2003, 
ApJ, 597, 210
\bibitem[Gavazzi et al. (2004)]{gavazzi04} 
Gavazzi G., Zaccardo A., Sanvito G., Boselli A., Bonfanti, C. 2004, A\&A, 417, 499
\bibitem[Grebenev et al. (1995)]{grebenev95} 
Grebenev S. A., Forman W., Jones C., Murray S., 1995, ApJ, 445, 607
\bibitem[Godwin \& Peach (1982)]{gp82} 
Godwin J. G., Peach J. V. 1982, MNRAS, 200, 733
\bibitem[Gratton et al. (2001)]{gratton01} 
Gratton R. G., Bonifacio P., Bragaglia A. et al., 2001, A\&A, 369, 87
\bibitem[Guenther (1994)]{guenther94} 
Guenther D. B., 1994, ApJ, 422, 400
\bibitem[Guzm\'an et al. (1992)]{guzman92} 
Guzm\'an R., Lucey J. R., Carter D., Terlevich R. J., 1992, MNRAS, 257, 187
\bibitem[Hempel \& Kissler-Pating (2004)]{hk04} 
Hempel M., Kissler-Patig M., 2004, A\&A, 428, 459
\bibitem[James et al. (2006)]{james06} 
James P. A., Salaris M., Davies J. I., Phillipps S., Cassisi S., 2006, MNRAS, 367, 339
\bibitem[Katz \& White (1993)]{kw93} 
Katz N., White S. W. D., 1993, ApJ, 412, 455
\bibitem[Kennicutt (1998)]{kennicutt98} 
Kennicutt R. C. 1998, ARA\&A, 36, 189
\bibitem[Kewley et al. (2006)]{kewley06} 
Kewley L. J., Geller M. J.,  Barton E. J. 2006, AJ, 131, 2004
\bibitem[Kewley et al. (2010)]{kewley10} 
Kewley L. J., Rupke D., Jabran Z. H., Geller M. J., Barton E. J., 2010, ApJ, 721L, 48 
\bibitem[Korn et al. (2006)]{korn06} 
Korn A. J., Grundahl F., Richard O., Barklem P. S., Mashonkina L., Collet R., Piskunov N., 
Gustafsson B. 2006, Nat, 442, 657 
\bibitem[Lane et al. (2007)]{lane07} 
Lane K. P., Gray M. E., Arag\'on-Salamanca A., Wolf C., Meisenheimer K., 2007, MNRAS, 378, 716
\bibitem[Lee et al. (2007)]{lee07} 
Lee H., Worthey G., Trager S. C., Faber S. M., 2007, ApJ, 664, 215
\bibitem[Lewis et al. (2002)]{lewis02} 
Lewis I., et al., 2002, MNRAS, 334, 673
\bibitem[Kroupa (2001)]{kroupa01} 
Kroupa P., 2001, MNRAS, 322, 231
\bibitem[MacArthur et al. (2004)]{macarthur04} 
MacArthur L. A., Courteau S., Bell E., Holtzman J., 2004, ApJS, 152, 175
\bibitem[Mihos \& Hernquist (1996)]{mh96} 
Mihos, J. C., \& Hernquist, L. 1996, ApJ, 464, 641
\bibitem[Michel-Dansac et al. (2008)]{michel-dansac08} 
Michel-Dansac L., Lambas D. G., Alonso M. S., Tissera P., 2008, MNRAS, 386, 82
\bibitem[Montuori et al. (2010)]{montuori10} 
Montuori M., Di Matteo P., Lehnert M.~D., Combes F., Semelin B., 2010, A\&A, 
submitted (arXiv:1003.1374)
\bibitem[Moss et al. (1998)]{moss98} 
Moss C., Whittle M., Pesce J. E., 1998, MNRAS, 300, 205
\bibitem[Moss \& Whittle (2000)]{mw00} 
Moss C., Whittle M., 2000, MNRAS, 317, 667
\bibitem[Moss \& Whittle (2005)]{mw05} 
Moss C., Whittle M., 2005, MNRAS, 357, 1337
\bibitem[Moss (2006)]{moss06} 
Moss C., 2006, MNRAS, 373, 167
\bibitem[Nelan et al. (2005)]{nelan05} 
Nelan J. E., Smith R. J., Hudson M. J., Wegner G. A., Lucey J. R., Moore S. A. W., Quinney S. J., 
Suntzeff N. B., 2005, ApJ, 632, 137
\bibitem[Odell et al. (2002)]{odell02} 
Odell A. P., Schombert J., Rakos K., 2002, AJ, 124, 3061
\bibitem[Peeples  et al. (2009)]{peeples09} 
Peeples M. S., Pogge R. W., Stanek K. Z., 2009, ApJ, 695, 259
\bibitem[Peletier et al. (1990)]{peletier90} 
Peletier R. F., Valentijn E. A., Jameson R. F., 1990, A\&A, 233, 62
\bibitem[Peletier \& Balcells  (1996)]{pb96} 
Peletier R. F., Balcells M., 1996, AJ, 111, 2238
\bibitem[Pietrinferni et al. (2004)]{pietrinferni04} 
Pietrinferni A., Cassisi S., Salaris M., Castelli F., 2004, ApJ, 612, 168
\bibitem[Poggianti (1997)]{poggianti97} 
Poggianti B., 1997, A\&AS, 122, 399
\bibitem[Puzia et al. (2002)]{puzia02} 
Puzia T. H., Zepf S. E., Kissler-Patig M., Hilker M., Minniti D., Goudfrooij P., 2002, A\&A, 391, 453
\bibitem[Rokas et al. (1996)]{rakos96} 
Rakos K., Maindl T. I., Schombert J., 1996, ApJ, 466, 122
\bibitem[Rakos \& Schombert (2005)]{rs05} 
Rakos K., Schombert J., 2005, AJ, 130, 1002
\bibitem[Rokas et al. (2007)]{rakos07} 
Rakos, K., Schombert, J., Odell, A., 2007, ApJ, 658, 929
\bibitem[Rokas et al. (2008)]{rakos08} 
Rakos, K., Schombert, J., Odell, A., 2008, ApJ, 677, 1019
\bibitem[Rose (1985)]{rose85} 
Rose J. A., 1985, AJ, 90, 1927
\bibitem[Rose et al. (1994)]{rose94} 
Rose J. A., Bower R. G., Caldwell N., Ellis R. S., Sharples R. M., Teague P., 1994, AJ, 108, 2054
\bibitem[Rupke et al. (2008)]{rupke08} 
Rupke D. S. N., Veilleux S., Baker A. J., 2008, ApJ, 674, 172
\bibitem[Rupke et al. (2010)]{rupke10} 
Rupke D. S. N.,  Kewley L. S., BArnes J. E., 2010, ApJ Letters, 710, 156
\bibitem[S\'anchez-Bl\'azquez et al. (2006)]{sanchez-blazquez06} 
S\'anchez-Bl\'azquez P., Gorgas J., Cardiel N., Gonz\'alez J. J., 2006, A\&A, 457, 809
\bibitem[Sakai et al. (2002)]{sakai02} 
Sakai S., Kennicutt R. C., van der Hulst J. M., Moss C., 2002, ApJ, 578, 842
\bibitem[Schlegel et al. (1998)]{schlegel98} 
Schlegel D. J., Finkbeiner D. P., Davis M., 1998, ApJ, 500, 525
\bibitem[Schiavon (2007)]{schiavon07} 
Schiavon R. P., 2007, ApJS, 171, 146
\bibitem[Schombert \& Rakos (2009a)]{sr09a} 
Schombert J., Rakos K., 2009a, AJ, 137, 528
\bibitem[Schombert \& Rakos (2009b)]{sr09b} 
Schombert J., Rakos K., 2009b, ApJ, 699, 1530 
\bibitem[Smail et al. (2001)]{smail01} 
Smail I., Kuntschner H., Kodama T., Smith G. P., Packham C., Fruchter A. S., Hook R. N., 2001, 
MNRAS, 323, 839
\bibitem[Smith et al. (2006)]{smith06} 
Smith R. J., Hudson M. J., Lucey J. R., Nelan J. E., Wegner G. A., 2006, MNRAS, 369, 1419
\bibitem[Smith et al. (2008)]{smith08} 
Smith R. J., Marzke, R. O., Hornschemeier A. E., Bridges, T. J., Hudson M. J., Miller N. A., 
Lucey J. R., V\'azquez G. A., Carter, D. 2008, MNRAS, 386, 96
\bibitem[Struble \& Rood (1999)]{sr99} 
Struble M. F., Rood H. J., 1999, ApJS, 125, 35
\bibitem[Sun \& Murray (2002)]{sm02} 
Sun M., Murray S. S. 2002, ApJ, 576, 708
\bibitem[Taylor et al. (2005)]{taylor05} 
Taylor V. A., Jansen R. A., Windhorst R. A., Odewahn S. C., Hibbard J. E., 2005, ApJ, 630, 784
\bibitem[Terlevich et al. (2001)]{terlevich01} Terlevich, A. I., Caldwell, N., Bower, R. G. 2001, 
MNRAS, 326, 1547
\bibitem[Terlevich \& Forbes (2002)]{tf02} 
Terlevich A. I., Forbes D. A. 2002, MNRAS, 330, 547
\bibitem[Thomas et al. (2003)]{thomas03} 
Thomas D., Maraston C., Bender R. 2003, MNRAS, 339, 897
\bibitem[Thomas et al. (2005)]{thomas05} 
Thomas D., Maraston C., Bender R., Mendes de Oliveira C., 2005, ApJ, 621, 673
\bibitem[Vazdekis (1999)]{vazdekis99} 
Vazdekis A., 1999, ApJ, 513, 224
\bibitem[West et al. (1991)]{west91} 
West M. J., Villumsen J. V., Dekel A., 1991, ApJ, 369, 287
\bibitem[Worthey (1994)]{worthey94} 
Worthey G., 1994, ApJS, 95, 107
\bibitem[Worthey \& Ottoviani (1997)]{wo97} 
Worthey G., Ottoviani D. L., 1997, ApJS, 111, 377
\bibitem[Zaritsky et al. (1994)]{zaritsky94} 
Zaritsky D., Kennicutt R. C., Jr., \& Huchra J. P. 1994, ApJ, 420, 87


\end{thebibliography}
\end{document}